\documentclass[preprints,article,submit,pdftex,moreauthors]{Definitions/mdpi} 
\firstpage{1} 
\pubvolume{1}
\issuenum{1}
\articlenumber{0}
\pubyear{2025}
\copyrightyear{2025}
\externaleditor{Firstname Lastname} 
\datereceived{2 April 2025} 
\daterevised{27 May 2025} 
\dateaccepted{24 June 2025} 
\datepublished{ }

\newcommand{\kms}{km s$^{-1}$}
\usepackage{cancel}
\usepackage{pdfpages}
\usepackage{lineno}

\hreflink{https://doi.org/} 

\Title{Two Cases of Non-Radial Filament Eruption and Associated CME Deflection}  

\TitleCitation{Two Cases of Non-Radial Filament Eruption and Associated CME Deflection}

\Author{Kostadinka Koleva 
 $^{1,}$*$^{,\dagger}$\orcidA{}, Ramesh Chandra $^{2}$\orcidB{}, Pooja Devi $^{2}$\orcidC{}, Peter Duchlev $^{1}$  and Momchil Dechev $^{3}$\orcidE{}}

\AuthorNames{Kostadinka Koleva, Ramesh Chandra, Pooja Devi, Peter Duchlev  and Momchil Dechev}

\isAPAStyle{%
       \AuthorCitation{Koleva, K.; Chandra, R.; Devi, P.; Duchlev, P.; Dechev, M.}
         }{%
        \isChicagoStyle{%
        \AuthorCitation{Koleva, K., Chandra, R., Devi, P., Duchlev, P.\& Dechev, M.}
        }{
        \AuthorCitation{Koleva, K.; Chandra, R.; Devi, P.; Duchlev, P.; Dechev, M.}
        }
}

\address{%
$^{1}$ \quad Space Research and Technology Institute, Bulgarian Academy of Sciences, 1113 Sofia, Bulgaria; kkoleva@space.bas.bg (K.K.) pduchlev@nao-rozhen.org (Peter Duchlev) 
\\
$^{2}$ \quad Department of Physics, DSB Campus, Kumaun University, Nainital 263001, India;\linebreak rchandra.ntl@gmail.com (R.C.); setiapooja.ps@gmail.com (P.D.)
\\

$^{3}$ \quad Institute of Astronomy and NAO, Bulgarian Academy of Sciences, 1784 Sofia, Bulgaria; mdechev@nao-rozhen.org}

\corres{Correspondence: kkoleva@space.bas.bg}

\abstract{The purpose of this paper is to analyze the multi-wavelength and multi-instrument observations of two quiescent filament eruptions  as well as the deflection of associated CMEs from the radial direction.  The events occurred on 18 October 2017 and 9 May 2021, respectively, in the southern solar hemisphere. Both of them and associated flares were registered by the Atmospheric Imaging Assembly (AIA) aboard the Solar Dynamics Observatory (SDO) and the Solar Terrestrial Relations Observatory--
Ahead (STEREO A) Observatory in different EUV wavebands. 
Using data from STEREO A COR1 and COR2 instruments and the Large Angle and Spectrometric Coronagraph (LASCO) onboard the Solar and Heliospheric Observatory (SOHO), we investigated morphology and kinematics of the eruptions and the latitudinal offset of the related CMEs with respect to the erupting filaments. Our observations provide the evidence that the two filament eruptions were highly non-radial. The observed deviations are attributed to the presence of low-latitude coronal holes.}


\begin{document}




\section{Introduction}
Solar prominences (also known as filaments when viewed against the solar disk) are among the most studied phenomena in the solar atmosphere. 
They are dense, cool structures that stretch outward into the solar corona and are frequently shaped like a loop. Numerous studies, including \citep{book1,labrosse,macay,parenti,gibson}, describe their nature.
In some circumstances, the prominences can lose their stability and erupt.  The type of eruption can be classified based on the relation between the prominence mass and the corresponding supporting magnetic structure \citep{gilbert,schr}. According to observations, three different types of prominence (filament) eruptions can be distinguished: full, partial, and failed (confined).
In cases of full eruption, the entire magnetic structure and the pre-eruptive prominence material are observed to eject into the heliosphere. Partial eruptions occur when the magnetic structure itself partially escapes with some or none of the filament mass.
The escaped filament material (or its part in case of partial eruption) is seen in the coronagraph observations as part of the Coronal Mass Ejection (CME) structure. 
An eruption is considered failed (confined) if neither the filament mass nor the supporting magnetic structure is able to escape the solar gravitational field. The kinematic evolution of the filament/prominence eruption usually shows two phases: an initial slow-rise phase, during which the filament gradually ascends with very low acceleration, followed by a sudden transition to a fast-acceleration phase~\citep{st2007, st2011}.

Observations show that during their eruption, prominences can occasionally deviate from the radial direction
~\citep{gopalTh, mccauley, devi}.
The study of \citep{williams}, for instance, noted that the flux rope's apex deviated from its radial direction in the direction of the weakest overlaying field. In the case of CMEs  triggered by an emerging flux, flux emergence away from the neutral line results in a non-radial eruption, according to the 2D magnetohydrodynamic (MHD) numerical simulations of \citep{chenshibata}.

Eruptive prominences (EPs) are frequently associated and physically related to CMEs and flares \citep{munro, stcyr, lin} 
 and some of them are followed by two ribbon flares \citep{choudhary, chandra}.
The erupted prominence material appears as a bright core of the CME structure in white-light coronagraph images \citep{chen, vourlidas, schmieder}.
 It has been observed that CMEs rotate, altering their orientation and deflecting from a purely radial trajectory. According to observations, extreme deflections can occur near the Sun \citep{gui, isavnin}, and the deflection motion can extend into interplanetary space \citep{lugaz, wang1}. Observational and numerical studies have shown that CMEs can deflect by ten or more degrees from the source location.
Both the prominences and the CMEs exhibited nonradial motions, and therefore, the CME deflects as a whole.

One of the main forces influencing space weather on the Earth is the propagation of CMEs. The arrival time at 1 AU can be altered by more than eight hours if the CME's direction varies by more than 10 degrees. 
Therefore, to comprehend the geoeffectiveness of solar eruptions, it can be essential to investigate the deflection of filament eruption and CMEs. 
The ultimate goal is to improve the ability to predict the arrival timing, speed, magnetic field, and geomagnetic effects of CMEs.

In this paper, we examined the morphology, kinematics, and related CME deflections of two filament eruptions, which occurred on 18 October 2017 and 9 May 2021, respectively.
%
The used data and methodology are introduced in Section~\ref{sec2}.  The results obtained from the study are presented in Section~\ref{sec3} and discussed in Section~\ref{sec4}. 

\section{Data and Methods of Analysis}\label{sec2}

For our current purposes, we used images taken with 1 min cadence in the He II 304 \AA` passband of the Atmospheric Imaging Assembly [AIA;\citep{lemen}] onboard the Solar Dynamics Observatory [SDO;\citep{pesnell}]. The AIA monitors the solar corona in  seven Extreme Ultra-Violet (EUV) and three Ultra-Violet (UV) channels  with a spatial resolution of $\sim 1.5^{\prime \prime}$. In this study we used data with 1 min cadence.
We also analyzed data from the Extreme Ultraviolet Imager (EUVI) onboard the Solar Terrestrial Relations Observatory [STEREO;\citep{kaiser}] Ahead (A) spacecraft to investigate the filament eruptions behind the limb and related flare for the event of 2017 October 18. 
The data used were taken with 10 min cadence in the He II 304 \AA\ and Fe XII 195 \AA\ channels. Furthermore, data from the COR2 outer coronagraph and the COR1 inner coronagraph \citep{thompson} on the STEREO A spacecraft were examined. 

Images obtained by the Large Angle and Spectrometric Coronagraph (LASCO) onboard the Solar and Heliospheric Observatory [SOHO;\citep{brueckner}]
were analyzed to study the associated CME and its non-radial motion. LASCO has two working coronagraphs, namely, C2 and C3, that observe the Sun in white-light from 2.5 to 30 R$_\odot$.

To investigate the kinematics of the events studied, we performed a time--distance analysis. This technique is based on investigating how plasma material moves along an artificial slit.
We investigated the non-radial offsets of the related CMEs, examining the variations of their position angles (PAs) over time. The PAs are measured counterclockwise from the solar North in degrees. We used the PA of the leading edge of the CME during its propagation in the plane of the sky. To determine the CME position angle, we used the online CME Catalog at the Coordinated Data Analysis Workshop (CDAW) Data Center~\citep{yashiro, gopal}, and the ``measuring'' tool therein. All the data analysis was performed by the solar software~(SSWIDL). 

\section{Results}\label{sec3}
\subsection{The 18 October 2017 Event}
The event from 2017 October 18 was observed by AIA/SDO as a prominence eruption (PE) at the east solar limb and was associated with a bright halo CME. The source region of the eruption was located behind the east solar limb in AIA observations. The event was observed as filament eruption near to the center via an EUVI instrument onboard the STEREO A observatory. 
At the time of observation, the separation angle of the STEREO A spacecraft with Earth was 125.617$^{\circ}$.

\subsubsection{Morphology}

\begin{figure}[H]
	\includegraphics[width=13.6 cm]{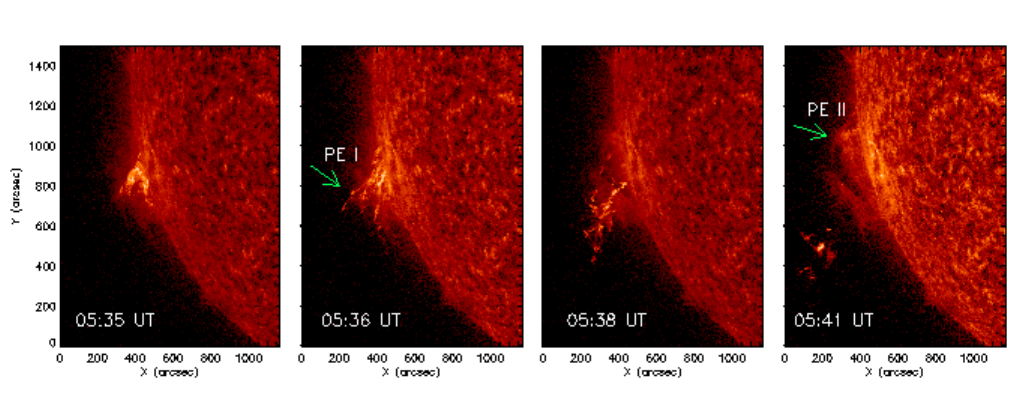}
\caption{Eruption 
 evolution as observed by AIA/SDO in He II 304 \AA\ channel.\label{mfig1}}
\end{figure}

In  the AIA field-of-view (FOV), prominence appears above the east limb as a compact bright structure. Two eruptions were observed at the same time. The first one, PE I, 
 was tracked from 05:30~UT to 05:44~UT.
The evolution of prominence eruptions is displayed in Figure~\ref{mfig1} in the AIA/SDO He II 304~\AA\ channel. In the figure, the top of PE I was visible, the legs of which were anchored behind the limb. A prominence cloud, PE II,was also visible below the eruptive structure. Around 05:32 UT, PE I erupted quickly, splitting in two branches.  In the AIA FOV, the eruption was visible up to $\sim$05:42~UT. Around 05:36~UT the prominence cloud was disturbed and started to erupt. From careful inspection of the movie we can infer that this second eruption reached up to 200$''$ from the limb around 05:57~UT. After that the prominence started to fall back on solar surface. Therefore, it is a failed eruption.
Since this eruption PE II was triggered by the first eruption, it is  a sympathetic failed eruption.

\vspace{-3pt}

\begin{figure}[H]
	\includegraphics[width=13.5 cm]{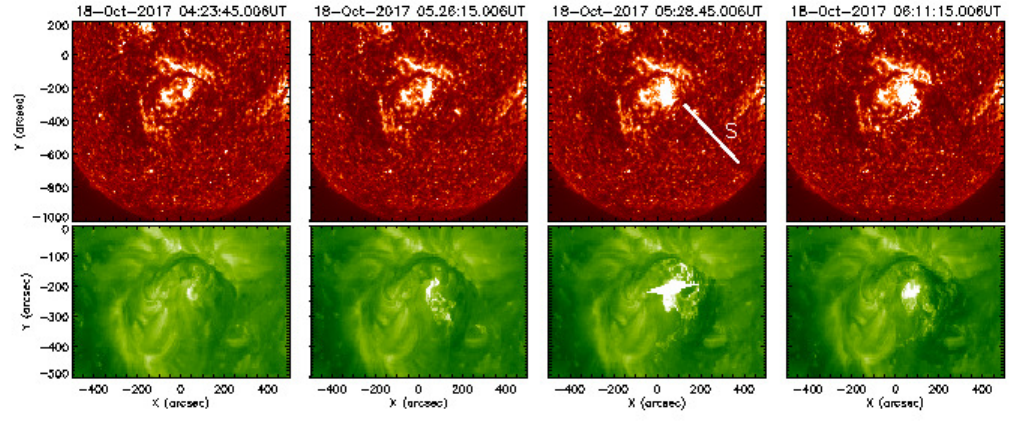}
	\caption{Filament eruption  observed by EUVI/STEREO A in He 304 \AA\ and 195 \AA\ channels. White line denotes slit position used to obtain time--distance plot shown in Figure~\ref{kfig5}.\label{mfig2}}
\end{figure}

\begin{figure}[H]
	\includegraphics[width=12.2cm]{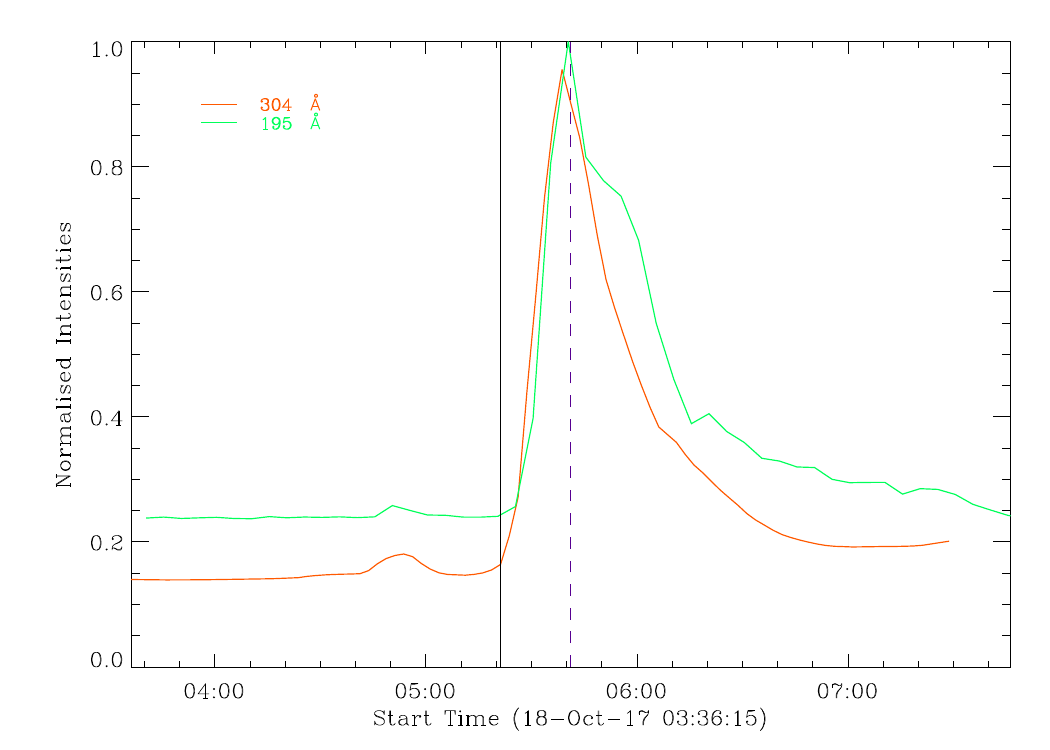}
	\caption{Light 
 curves of the flare region in 304 \AA\ (red line) and in 195 \AA\ (green line). The black line shows the flare start at 05:21~UT and the dotted purple line denotes the flare maximum at 05:41~UT.\label{mfig3}}
\end{figure}

Now let us discuss the eruption of filament visible in EUVI/STEREO--A images. In STEREO--A observations the eruption started $\sim$05:26 UT, which is about 6 min earlier than the AIA observations. The delay in the appearance of the AIA observations is due to its location, which is behind the east limb. The evolution of the eruption in STEREO--A observations is presented in Figure~\ref{mfig2} in~304~\AA~and~195~\AA~images. 
 Before the eruption, the filament appeared as a circular filament and its eruption was linked to a solar flare. Due to the location of eruption, we could not see the GOES observations of this flare. The related  flare began at approximately 05:20~UT close to the center of the circular filament and reached its maximum at 05:41~UT. We created the light curves of the flaring region in 304 \AA\ and 195 \AA, shown in Figure~\ref{mfig3}, in order to determine the exact flare time. From this curve, we estimated the flare onset time and it was found to be $\sim$05:21 UT.

\vspace{-3pt}
 \unskip

Part of the filament became active and started to rise slowly at around 04:20~UT. 
At about 05:20~UT, shortly after the flare start, the eruption entered its fast-rising phase. One of the filament legs separated from the surface at approximately 05:41~UT, when the flare was at its maximum, and the filament partially erupted. 
When this filament eruption reached about 200 arcsec from the origin site, it encountered another filament (PE II). As a result of this, this distant filament was disturbed and started to erupt. It erupted up to some distance, and after that the filament material dissipated. 
The falling back material cannot be clearly seen in AIA images.
\begin{figure}[H]
	\includegraphics[width=9 cm]{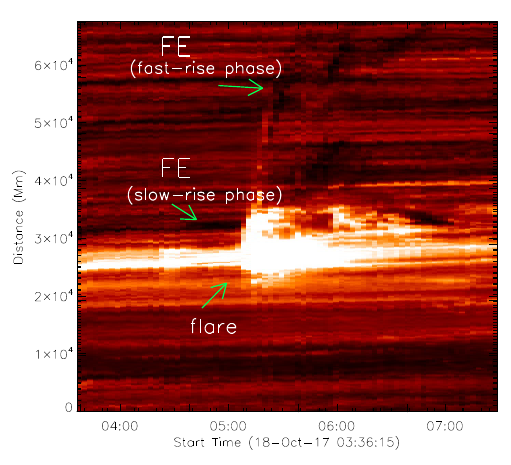}
	\caption{Time--distance plot taken in the EUVI/STEREO A 304 \AA\ channel, corresponding to the slice
shown through the white line in Figure~\ref{mfig2}.\label{kfig5}}
\end{figure}   

\subsubsection{Kinematics}
\begin{figure}[H]
		\subfloat[\centering]{\includegraphics[width=6.6cm]{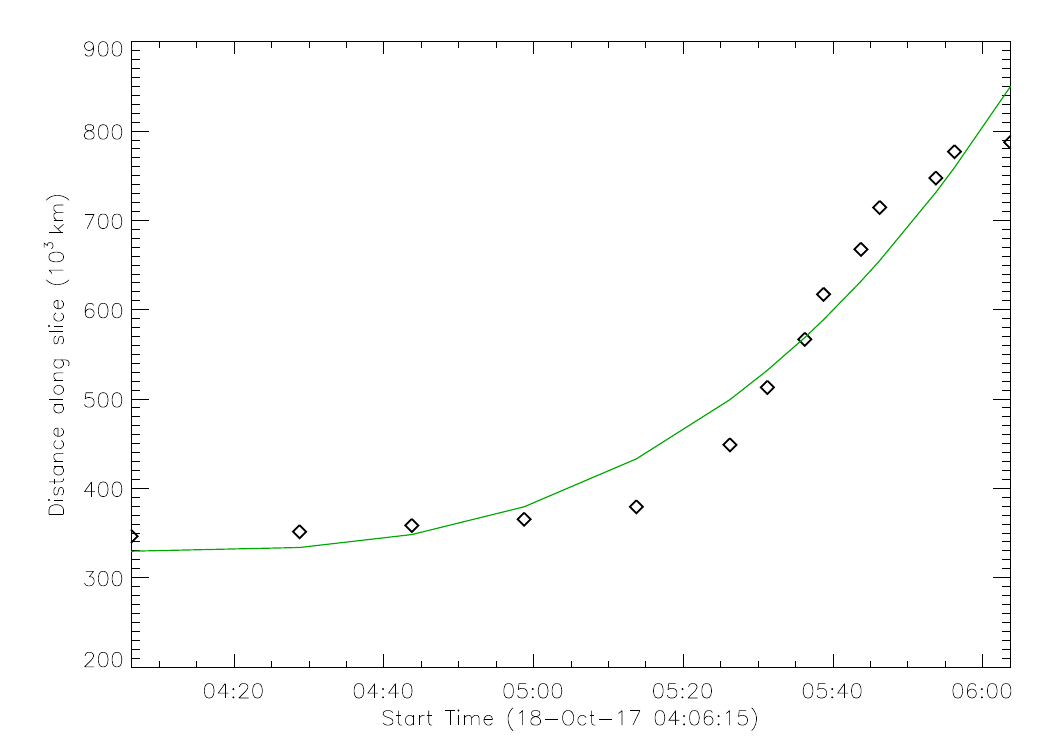}}
		\subfloat[\centering]{\includegraphics[width=6.6cm]{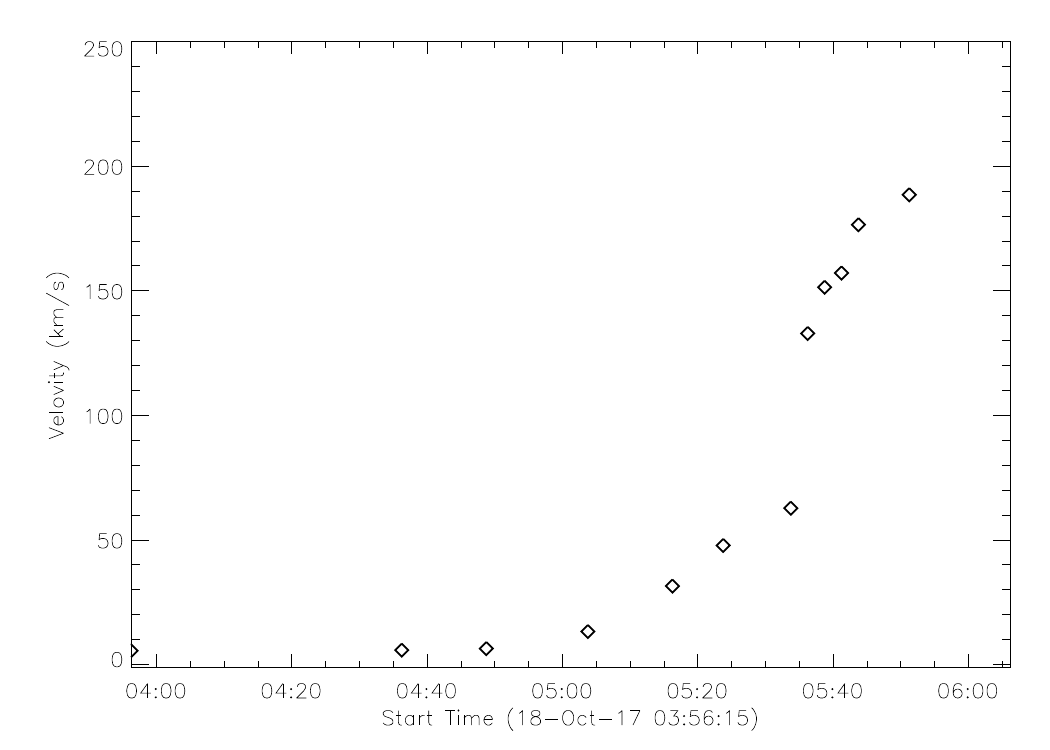}}\\
		
	\caption{(\textbf{a}) Time--distance plot of the filament eruption as seen from STEREO A FOV.  (\textbf{b}) Velocity changes during the eruption. The green solid line is the fitting curve to the data points along the slice.\label{kfig6}}
\end{figure}

We investigated the kinematics of the event using both the  observations, i.e., AIA/SDO and EUVI/STEREO--A. 
For the time--distance analysis, using the STEREO--A 304 \AA\ data, the selected artificial slit is shown in Figure \ref{mfig2} and the corresponding time--distance plot is presented in Figure \ref{kfig5}. From this time--distance plot, we obtained the distance--time data by tracing the leading edge of the erupting filament. The variation of erupting filament height with time is shown in Figure \ref{kfig6}a. Then we fit a second-order polynomial to the tracked data which is represented by the green curve in the figure.
The velocity of the filament is presented in Figure~\ref{kfig6}, in the right panel.
For hot channels, the uncertainty of measured heights should be two or three pixels \citep{cheng}. 
The filament began to slowly rise, moving at a speed from 5.5 \kms~ to 6.3 \kms. Until 05:20~UT, when the flare started, the changes in filament distance along the slice were insignificant. After that time, a sudden increase in height was observed. From 05:20~UT to 06:00~UT, when the filament loop left the EUVI FOV, the height along the slice changed from 379 $\times$ $10^3$ km to 787 $\times~10^3$ km and the speed increased from 6 to around 243 \kms~ with an average acceleration of about 140 m s$^{-2}$.



\vspace{-14pt}
\clearpage
 
\begin{figure}[H]
	\includegraphics[width=8.2 cm]{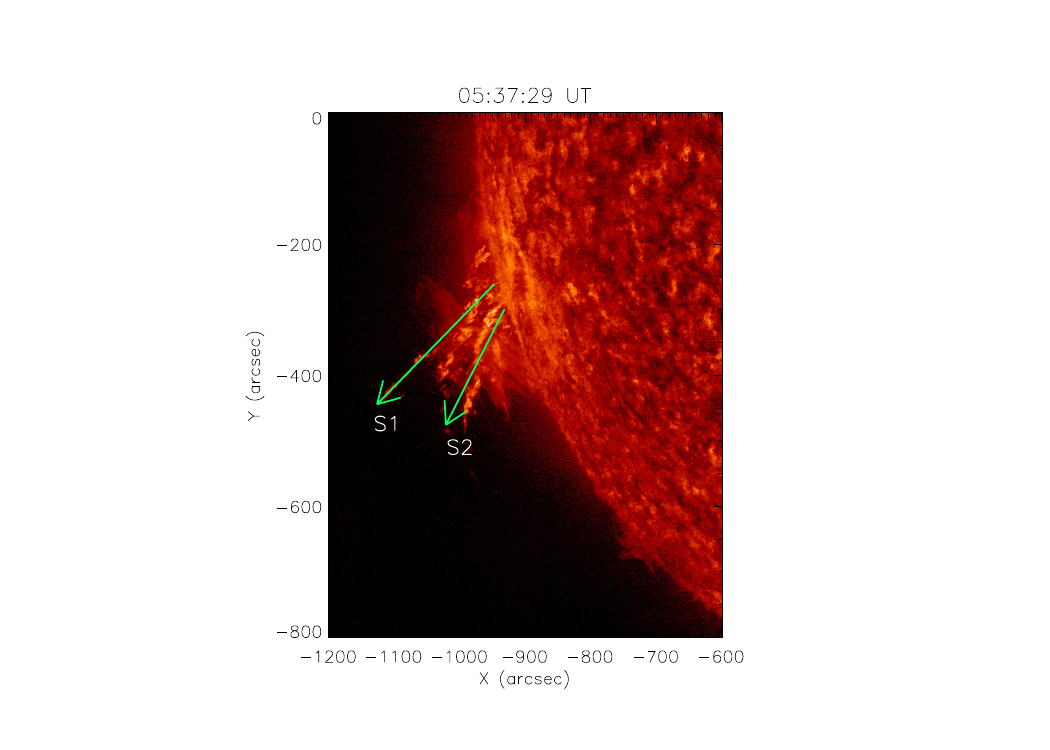}
	\caption{The slit position used to obtain time--distance plot in the AIA FOV. \label{kfig5_1}}
\end{figure}

\begin{figure}[H]
	\centering
        \subfloat[\centering]{\includegraphics[width=7.5 cm]{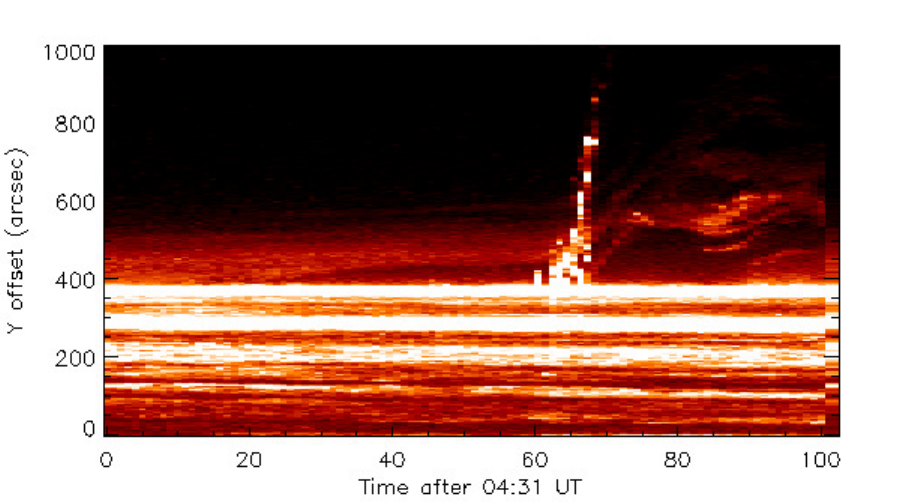}}
        \subfloat[\centering]{\includegraphics[width=5.5 cm]{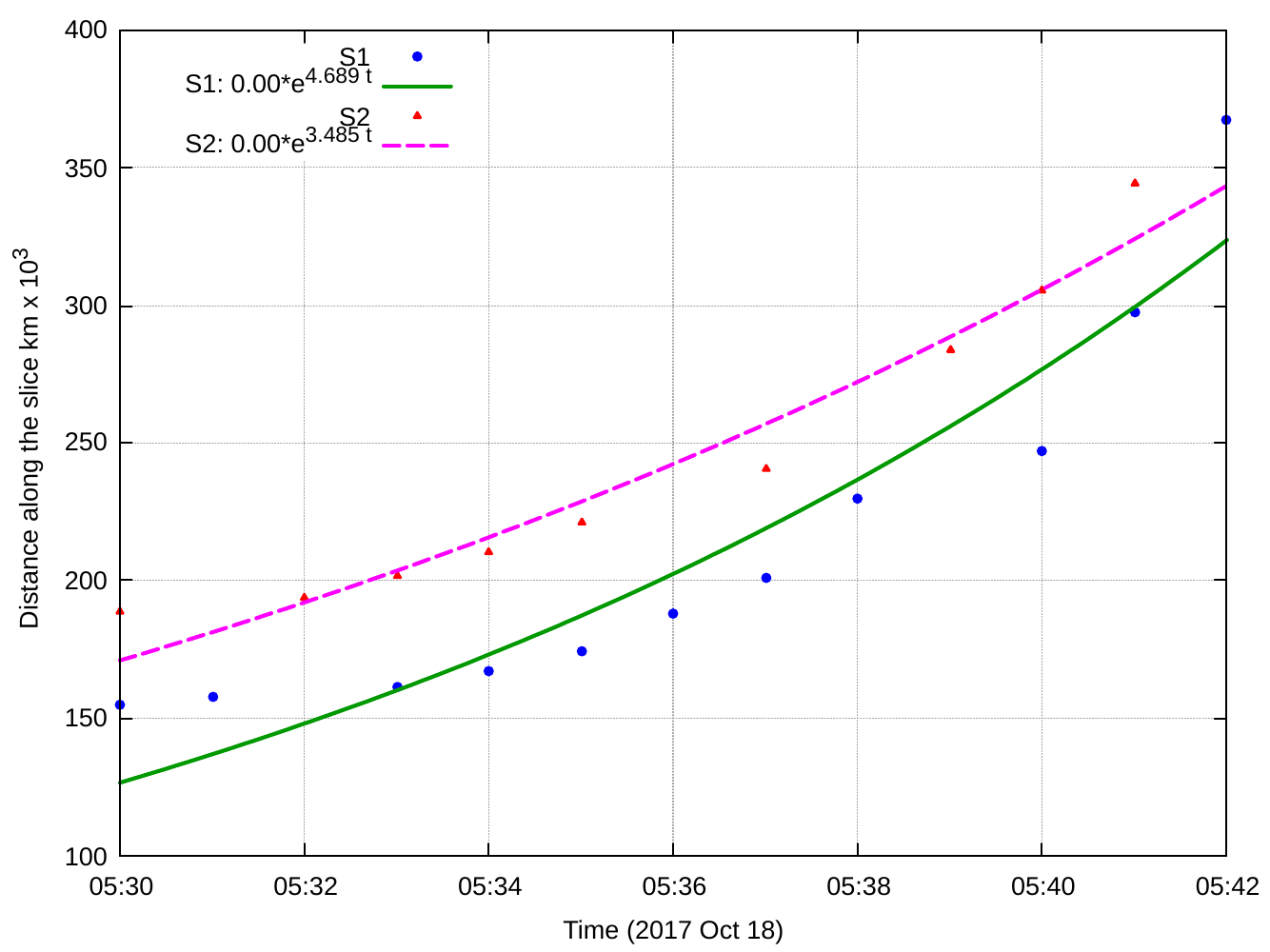}}\\
		
	\caption{(\textbf{a}) Time--distance plot taken in the AIA 304 \AA\ channel, corresponding to the slice S1. \\ (\textbf{b})  Time--distance plot of the prominence eruption along the slits S1 and S2. \label{kfig6_1}}
\end{figure}

In a similar way, we selected two slits, namely S1 and S2, for the kinematic study in the AIA FOV, which are shown in Figure~\ref{kfig5_1}. The time--distance plot for the S1 slice and the corresponding distance-time profile are shown in Figure~\ref{kfig6_1}a,b, respectively. When observed in AIA FOV, the eruption was already in progress. Therefore, the first slow-rise phase was not visible in the AIA images. The kinematic behavior of the two branches was quite similar. The prominence rises with increasing velocity of 261 \kms, and average acceleration of 68 ms$^{-2}$, and finally reached a height of about 368 $\times~10^3$ km.

\vspace{-6pt}
\vspace{-18pt}

\begin{figure}[H]

\begin{adjustwidth}{-\extralength}{0cm}
	\centering
	\subfloat[\centering]{\includegraphics[width=6.8cm]{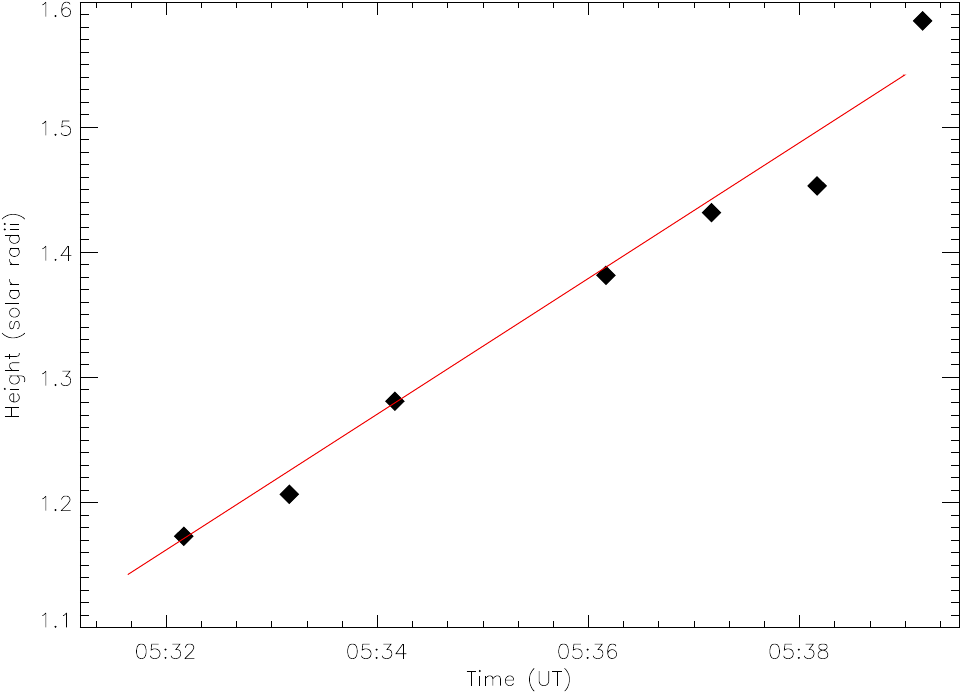}}\\\vspace{-12pt}
	\hfill
	\subfloat[\centering]{\includegraphics[width=7.2cm]{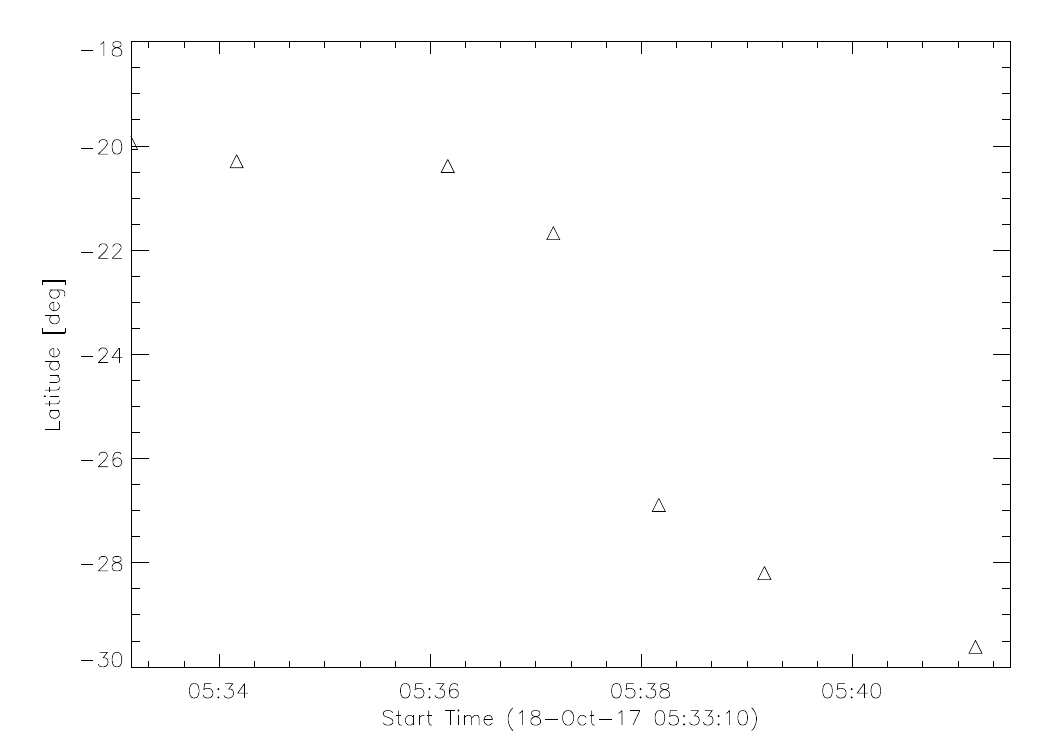}}
	\subfloat[\centering]{\includegraphics[width=7.2cm]{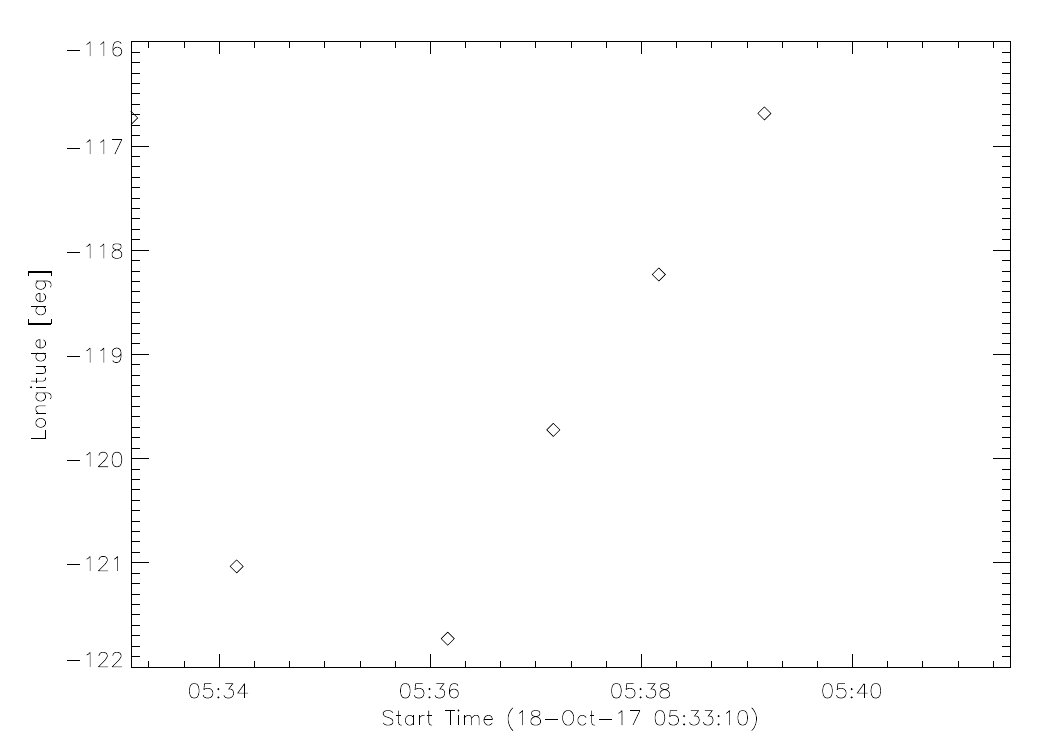}}
\end{adjustwidth}
		\vspace{-3pt}
	\caption{(\textbf{a}) Time--
 distance plot for the fast-rise phase of filament eruption from 18 October 2017.  (\textbf{b})~Evolution of the latitude  and longitude (\textbf{c}) of the top of the filament loop.\label{kfig8_new}} 
\end{figure}

We also used the \texttt{scc\_measure.pro} routine,  available  in the SSW package, to  obtain the location of the filament top in 3D, including its true height, latitude, and longitude. The result for part of the fast-rise eruption phase is presented in Figure~\ref{kfig8_new}. From 05:31~UT to around 05:40~UT the filament’s true height was changed from 1.1 $R_\odot$ to 1.6 $R_\odot$.
For the same time period, the latitudinal direction changes from $-20$ to $-30$ degrees, while the longitudinal direction moved  from $-121^\circ$ to around $-116.7^\circ$.

\vspace{-6pt}

\subsubsection{CME Deflection}

\begin{figure}[H]
	\centering
	\subfloat[\centering]{\includegraphics[width=5.7cm]{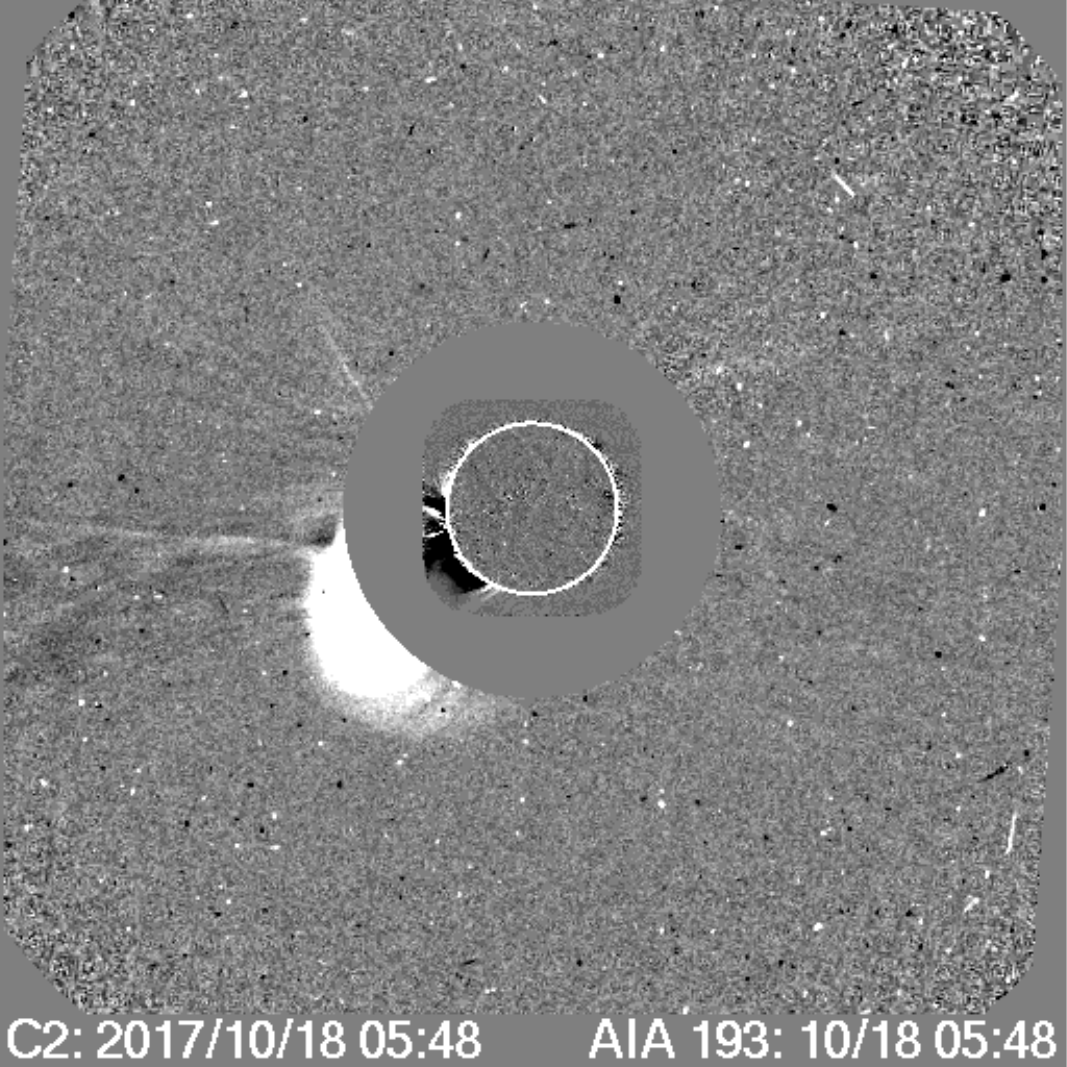}}
	\hfill
	\subfloat[\centering]{\includegraphics[width=5.7cm]{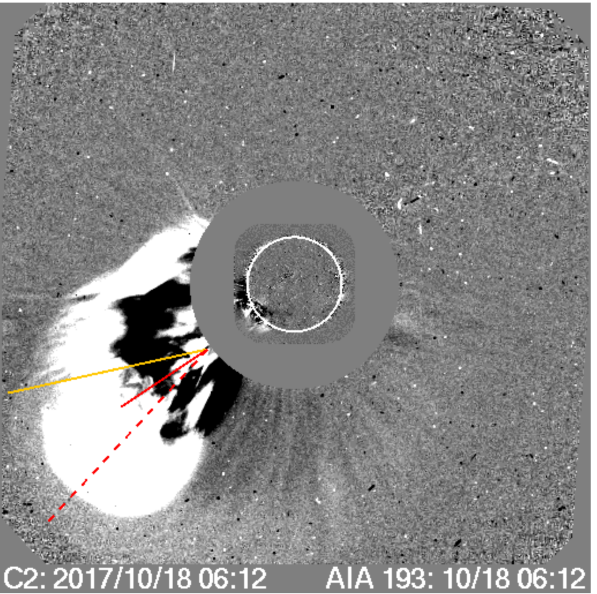}}\\
	
	\caption{(\textbf{a}) SOHO/LASCO/C2 white-light and AIA/SDO 193 \AA\ running differential composite image at 05:48 UT. (\textbf{b}) Image at 06:12 UT. Initial PE direction is indicated by a yellow line. The CME PAs at 05:48 UT and 06:12 UT are shown by solid and dashed red
		lines, respectively.\label{cmefig7}}
\end{figure}

The associated CME first appeared at the LASCO C2 FOV at 05:48~UT and had a linear speed of 1576 \kms. The PE started at a PA of $108^\circ$. During its propagation, a significant non-radial offset between the initial prominence location and the associated CME loop was observed. The CME PA at two separate times, namely at 06:12~UT and 05:48~UT, are shown in Figure \ref{cmefig7} with red dashed and solid lines, respectively. The yellow solid line in this figure represents the direction of PE in AIA FOV.

\vspace{-3pt}

\begin{figure}[H]
		\subfloat[\centering]{\includegraphics[width=6.8cm]{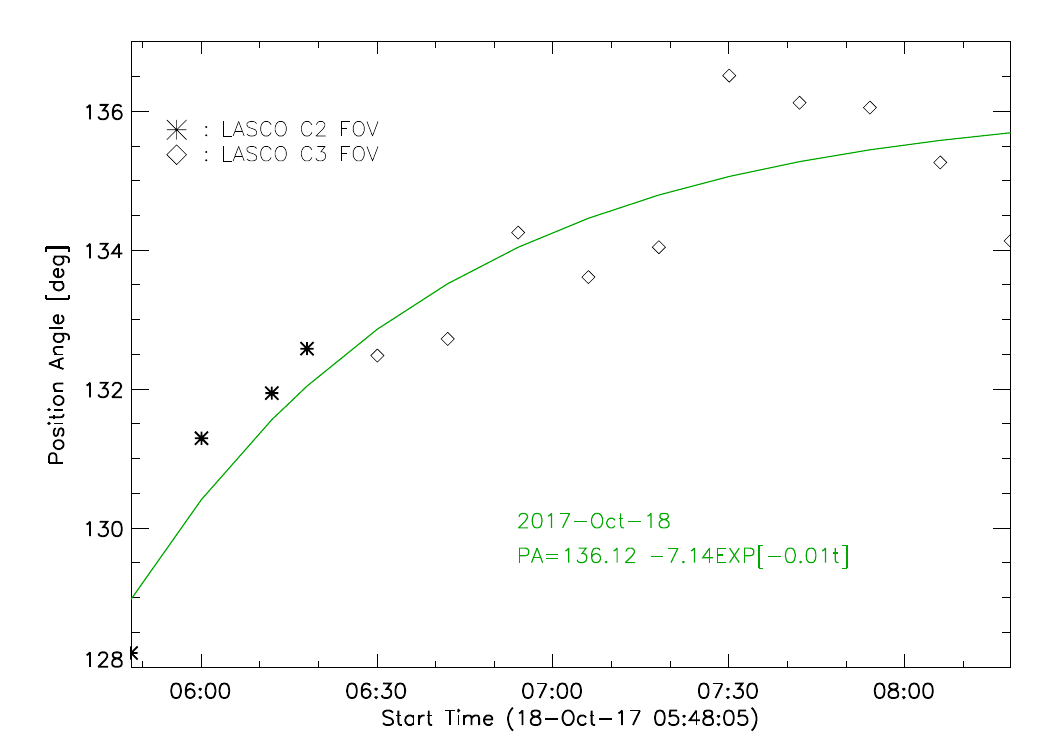}}
		\subfloat[\centering]{\includegraphics[width=6.8cm]{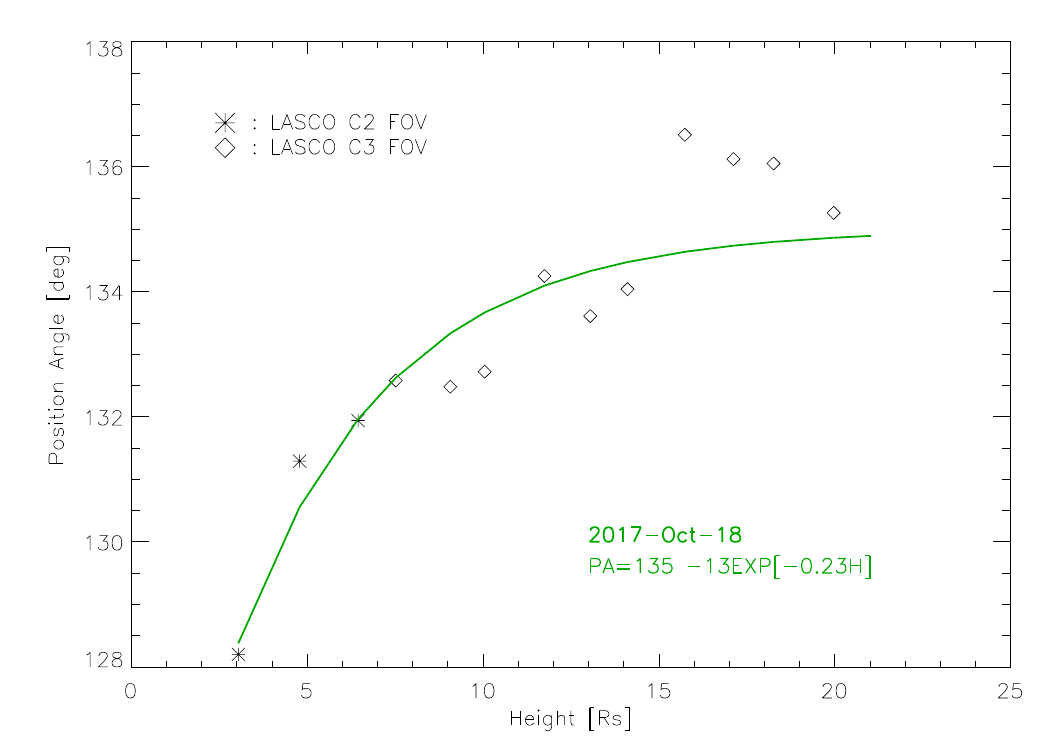}}\\
		
	\caption{(\textbf{a})  Variation of the position angle of the CME nose as a function of time. Stars and diamonds
		represent measurements, respectively, from C2 and C3 coronagraphs of SOHO/LASCO. (\textbf{b}) CME PA
		at various heliocentric distances. The solid line is the fit to the data points, showing that the non-radial
		motion gradually decreases at about \textbf{15} $R_\odot$ .\label{cmefig8}}
\end{figure}

We measured the position angle of the CME nose in the C2 and C3 LASCO coronagraphs, as described in Section~\ref{sec2}. 
At 08:24~UT the CME's nose PA was of about $137^\circ$ and the measured latitudinal deviation was $29^\circ$ to the pole. In Figure~\ref{cmefig8}, the variation of the PA of CME as a function of time (left) and heliocentric distances (right) is presented. The solid lines are the exponential fits to the data points. The non-radial motion gradually decreased and stopped at about 15 $R_\odot$. 
\vspace{-6pt}

\subsection{The 9 May 2021 Event}

The prominence eruption from 9 May 2021 was observed by the SDO observatory in the southern solar hemisphere between 9:30 and 11:48~UT. The eruption started at the heliographic coordinates of ($-26.87^\circ$), ($-13.88^\circ$), and PA $154^\circ$, close to the plage region and was accompanied by spreading ribbons at its base. The STEREO--A  spacecraft also recorded the eruption near the west limb.
The separation angle of STEREO--A from Earth at the time of observation was $52^\circ$. In STEREO--A FOV the eruption was associated with a CME.

\subsubsection{Morphology}
\begin{figure}[H]
	\includegraphics[width=13 cm]{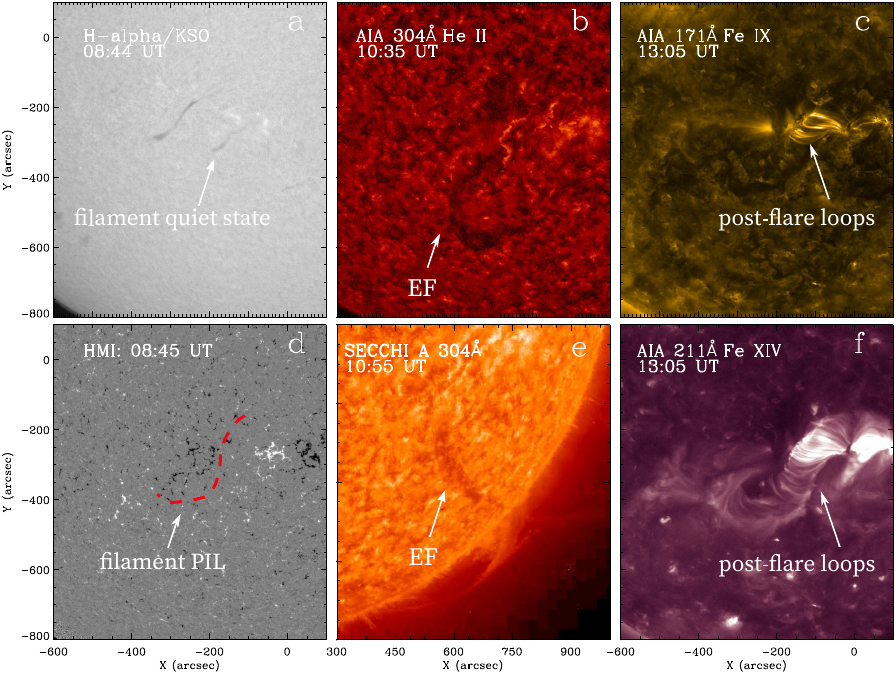}
	\caption{(\textbf{a}) Image of the event in H-alpha, showing the filament in the quiet state before the eruption.
(\textbf{b}) The EFs observed in AIA 304 \AA\ and (\textbf{c}) in AIA 171 \AA. (\textbf{d}) HMI magnetogram with over-plotted
the filament PIL position (red dotted line). (\textbf{e}) Filament eruption observed by EUVI/STEREO A
in 304 \AA. \textbf{(f)} Image in the AIA 211 \AA channels, showing the post-flare loops.\label{mayfig8}}
\end{figure}

Before the eruption, the filament represented a sigmoid filament and was located along the S-shaped magnetic polarity inversion line (PIL).
Figure~\ref{mayfig8} displays the event morphology in several pass bands. The filament's `S' shape was clearly seen in the Kanzelh\"{o}he Solar Observatory (KSO) H-alpha images that we used to trace it in the quiet state (Figure~\ref{mayfig8}a). 
Figure~\ref{mayfig8}d represents the HMI magnetogram, showing the line-of-sight photospheric magnetic fields at 08:45 UT. The S-shaped filament channel is traced by a red dotted~line. 

The filament in AIA/SDO FOV began to rise slowly at 09:30 UT and erupted at around 10:00~UT.
Following that, the filament eruption (FE) began its fast-rising phase.  Part of the filament material was ejected in the west direction during the eruption, while the major part deviated towards the east direction. Between 11:18~UT and 12:18~UT, the erupting motion was accompanied by flux rope rotation, and the top of the filament loop deviated toward the northeast. In Figure~\ref{mayfig8}b the FE in AIA FOV is shown at 10:35~UT, when the filament loop erupted radially with increasing velocity.
The filament eruption on the west limb, as observed by STEREO--A in the EUVI 304 \AA\ channel, is shown in Figure~\ref{mayfig8}e.

Following the onset of the eruption, flare ribbons were seen to form along the filament channels two sides, and post-flare loops developed above it. These events were well-observed in hotter AIA channels, such as 171 \AA\ (Figure~\ref{mayfig8}c) and 211 \AA\ (Figure~\ref{mayfig8}f).

\subsubsection{Kinematics}

\begin{figure}[H]
	\includegraphics[width=13 cm]{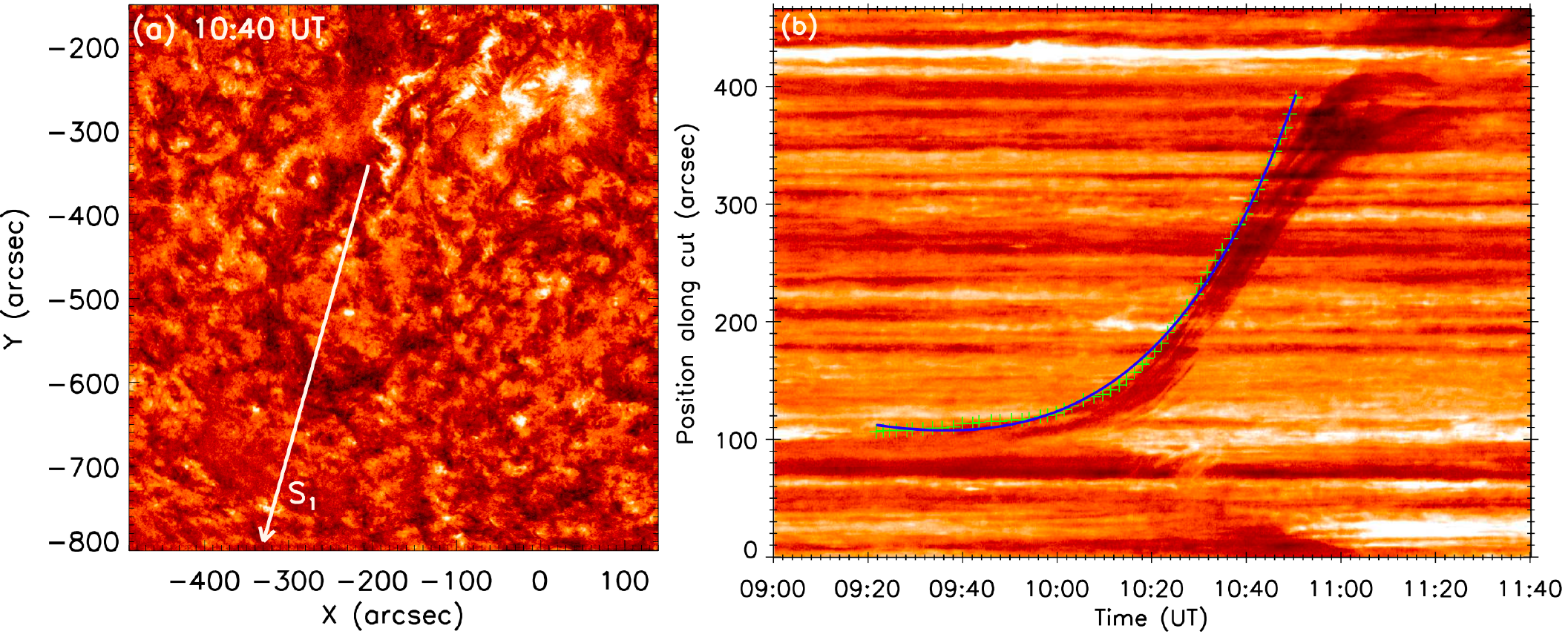}
	\caption{Slice position for the time--distance plot of the filament eruption in the AIA FOV (\textbf{a}) and the corresponding  time--distance plot along the slit S1 (\textbf{b}).\label{mayfig9}}
\end{figure}

We used the distance--time plot to investigate the eruption kinematics in the AIA FOV as well as in STEREO--A images.  Following the direction of filament material ejection in the AIA FOV, we have chosen the artificial slice (S1) shown in Figure~\ref{mayfig9}a). The corresponding time--distance plot is presented in Figure~\ref{mayfig9}b).
The green `plus' symbols in panel (b) in Figure~\ref{mayfig9} are the data points chosen from the time--distance plot and blue solid line is the fitting curve to these data points. The fitted function is taken from \citep{cheng}. 

Around 9:30~UT, filament started to erupt and the filament material began to rise slowly in the southeast direction. Two distinct eruptive phases were evident in the eruption: the slow-rise
phase, lasting between 09:30~UT and 09:58~UT, and the fast-rise phase after that time. From the first and second derivative of the fit of the distance--time curve, we defined the speed and acceleration of the eruption.
The eruption speed in the fast-rise phase varied from $\sim$ 20 \kms~ to 140 \kms. The maximum acceleration calculated in this time period was 70 m s$^{-2}$. After 11:48~UT, the filament loop started to deviate in a northeast direction.
 The eruption started at a position angle of about $153^\circ$. After deviation, the position angle of the center of the prominence loop was about $142^\circ$. 

\vspace{-3pt}
\begin{figure}[H]
	\includegraphics[width=11 cm]{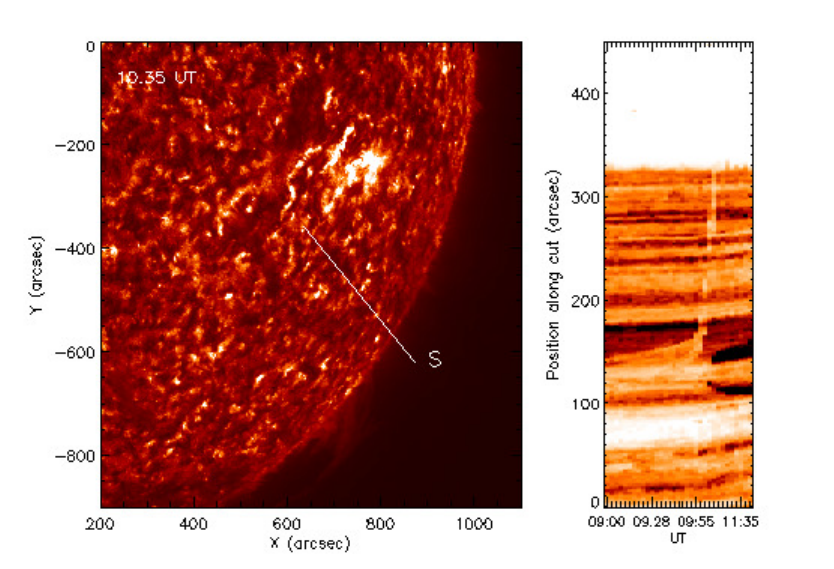}
	\caption{(\textbf{left}) Location of time slice selected for FE direction in EUVI/STEREO A
FOV. (\textbf{right}) Corresponding time--distance plot in reversed color table. \label{mayfig10}}
\end{figure}   

\begin{figure}[H]
	\includegraphics[width=8.5 cm]{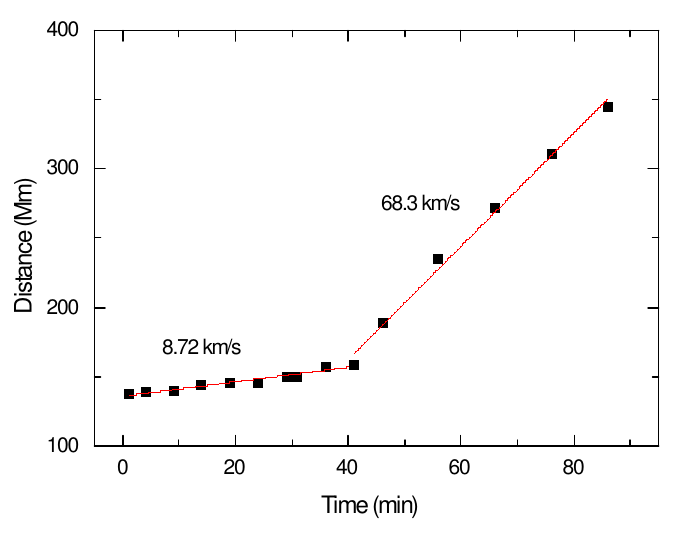}
	\caption{Time--distance profile of filament eruption for two phases of its evolution: slow-rise phase and fast-rise phase. Time is minutes after 09:00~UT.\label{mayfig11}}
\end{figure}

We used data from the EUVI instrument in the 304 \AA\ channel to analyze the event kinematics in the STEREO A field of view. Figure~\ref{mayfig10} illustrates the location of the selected slice along with the corresponding time--distance plot. The eruption within the EUVI FOV once again showed two separate phases that could be clearly distinguished by their velocities. Two linear fits to the data points were used to calculate the eruption speed (Figure~\ref{mayfig11}).
The slow rise motion of the filament was observed after 09:00~UT and lasted until 09:55~UT. The average estimated velocity for this phase is about 9 \kms.  During the second phase, the filament material continued to rise rapidly with an average speed of about 68 \kms. The filament material reaches a distance of about 320 Mm in the STEREO~FOV.
\vspace{-6pt}

\subsubsection{CME Deflection}

\begin{figure}[H]
	\includegraphics[width=9 cm]{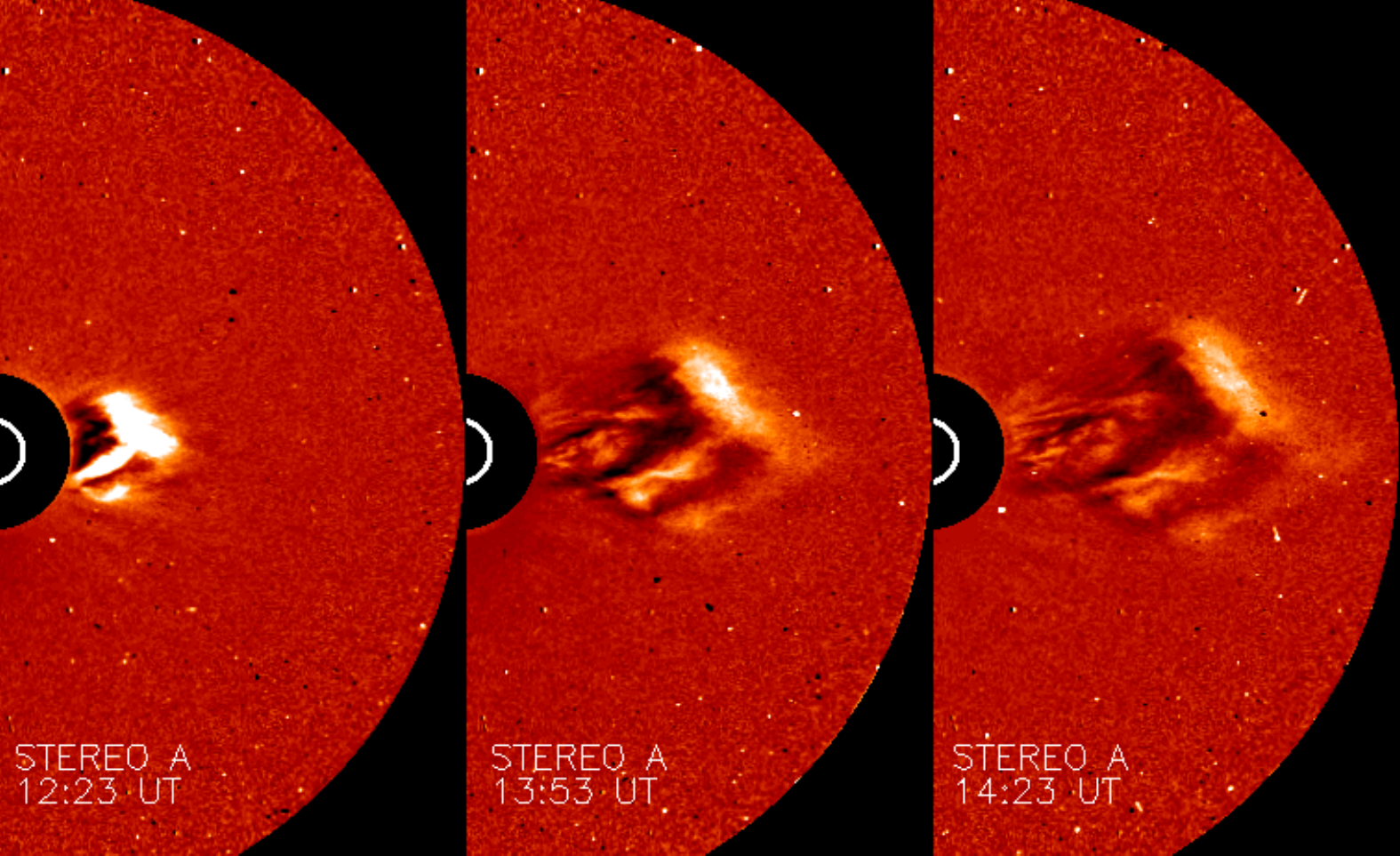}
	\caption{Propagation of  associated CME in STEREO~A/COR2 FOV.\label{fig4}}
\end{figure}

\begin{figure}[H]
		\subfloat[\centering]{\includegraphics[width=6.8cm]{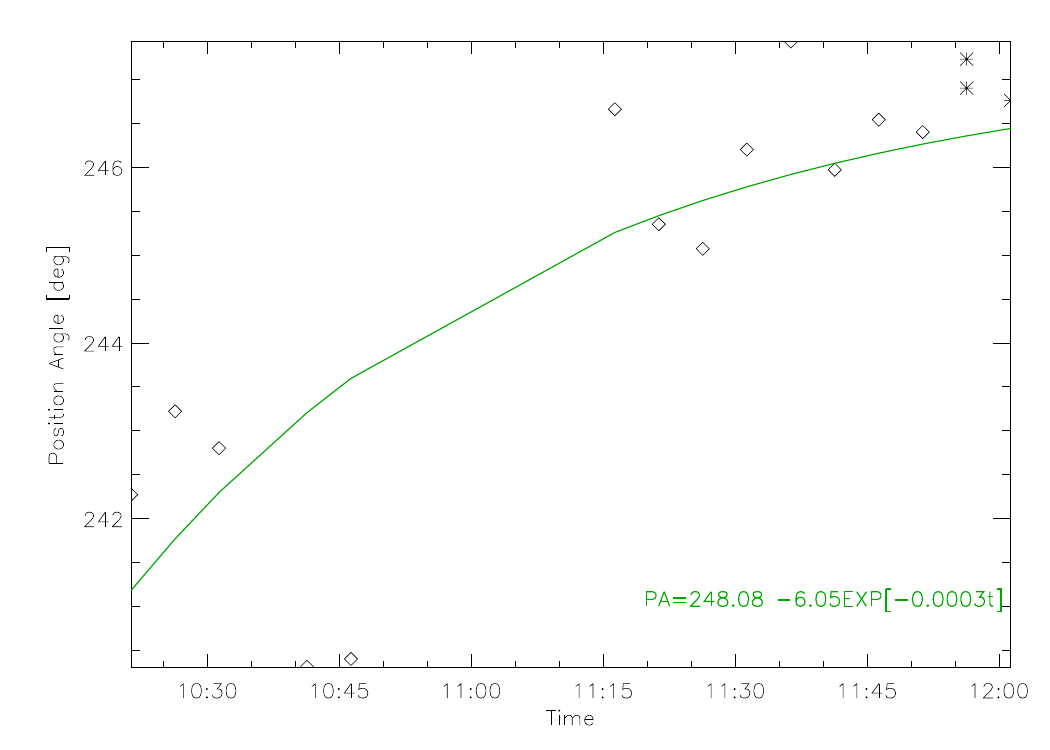}}
		\subfloat[\centering]{\includegraphics[width=6.8cm]{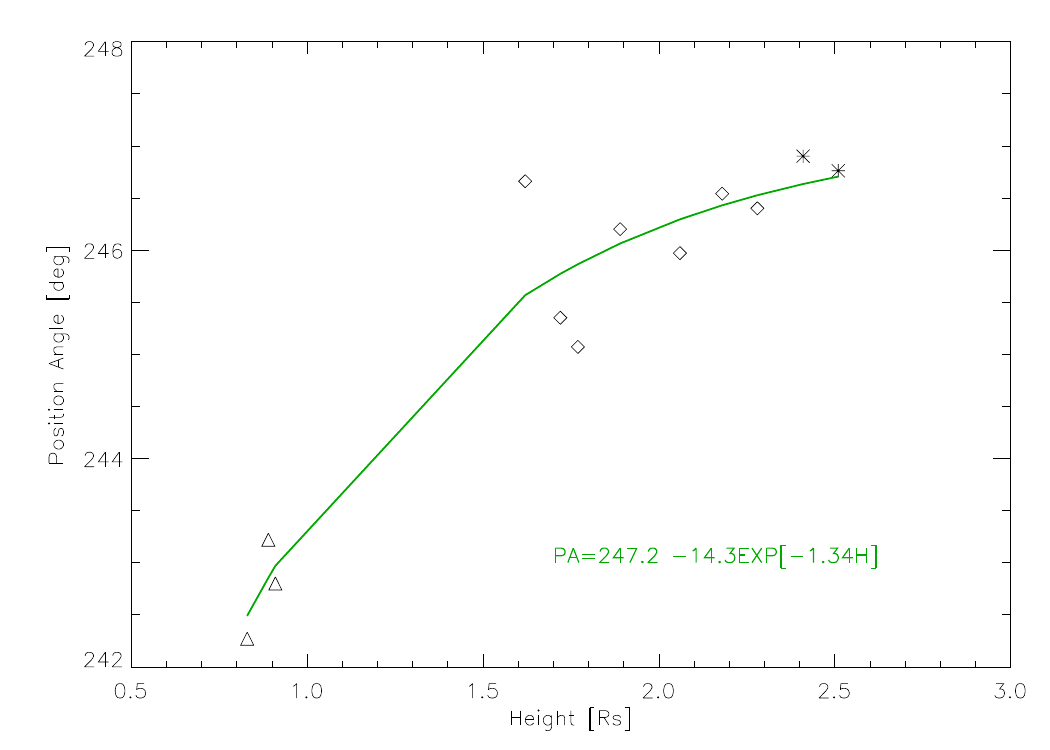}}\\
		
	\caption{Variation in position angle of CME core  as function of time and heliocentric distances: (\textbf{a})  PA vs. time variations. Diamonds and stars represent measurements from COR1 and COR2 coronagraphs of STEREO A, respectively. (\textbf{b}). Prominence PA at various heliocentric distances. Triangles, diamonds and stars represent measurements from disk (EUVI 195\AA), COR1, and COR2, respectively. Solid line is  fit to  data. \label{fig5}}
\end{figure}

The eruption was associated with CME, which was clearly visible in the FOV of the STEREO--A coronagraph (Figure~\ref{fig4}). However, the CME is not visible in LASCO observations. This could be due to the non-radial eruption of the filament. Initially, the major part of the erupted filament, as observed by SDO, deviated eastward  to the Earth direction, and was observed as a CME by COR 1 and COR 2 coronagraphs of STEREO A.
Between 10:21~UT and 12:01~UT, the CME spread throughout the COR1 FOV. During this time period, the PA changed from $241^\circ$ to $247^\circ$. 
We analyzed the variations of the CME's position angle as a function of time and heliocentric distances, and the results are presented in Figure~\ref{fig5}.

The non-radial motion stopped at 2.25 R$_\odot$. After that, the core continues its propagation in the C2 FOV without visible deflection (Figure~\ref{fig5}b).

\section{Discussion and Conclusions}\label{sec4}
We conducted two case studies of filament eruptions that took place on 18 October 2017 and 9 May 2021, respectively, in the southern solar hemisphere. Both events were associated with CMEs and solar flares. 
The data from SDO and STEREO--A observatories were used to analyze the kinematics and morphology of the events from two different observing~angles. 

The observational details of the eruptions and their kinematics are summarized in Table~\ref{tab1}.
\vspace{-3pt}
\begin{table}[H] 
\tablesize{\small}
\caption{Summary of the eruptions kinematics and the associated CMEs.\label{tab1}}
\begin{tabularx}{\textwidth}{CCCCC}
\toprule
&\textbf{18 October 2017}&	& \textbf{9 May 2021}	& \\
\midrule
&SDO FOV & 	STA FOV	& SDO FOV  &STA FOV\\
\midrule
Location &SE limb &$-$38, $-$223 & $-$138, $-$457 & SW limb\\
FE start &05:32~UT & 05:26~UT&09:30~UT &09:40~UT\\
FE velocity: & & & &\\
slow phase& -- & 5.84 
 \kms\ &20 \kms\ &8.72 \kms \\
fast phase 
& 261 \kms\ & 200 \kms &140 \kms\ & 68.3 \kms \\
CME& 05:48~UT &06:54 (COR2) & No CME associated & 10:21~UT\\
Latitudinal deviation & $29^\circ$ \textsuperscript{1} & - & $6^\circ$ \textsuperscript{2} &\\
Flare (start/maximum)& &05:20/05:41 & spreading ribbons&\\
\bottomrule
\end{tabularx}

\noindent{\footnotesize{\textsuperscript{1} Towards the pole.}}
\noindent{\footnotesize{\textsuperscript{2} Towards the equator.}}
\end{table}

\vspace{-6pt}
From the analysis of the eruptions' kinematic characteristics, it is evident that there are two distinct eruptive phases in both observed cases: a slow-rise and a fast-rise phase. 
The source region of the eruption from 2017 October 18 was observed by the STEREO--A observatory. Before the eruption, the filament had a circular shape and was linked to a solar flare. The filament partially erupted. The eruption velocities ranged from 0 to $\sim$ 6 \kms~ during the slow-rising phase and reached speeds higher than 200 km~s$^{-1}$ during the fast-rising phase.
The filament activation coincided with the start of the flare, while the fast-rising phase started when the flare reached its maximum phase.
As the filament slow-rising phrase and the pre-flare heating appear to have occurred simultaneously, we can speculate that because of the increasing temperature, the filament structure became less stable and started to rise slowly. 
Such thermal non-equilibrium was proposed by \citep{hood}, who found that the lack of thermal equilibrium can result in magnetic disturbances or particle acceleration. 
Due to the total loss of equilibrium, the part of the filament erupted outwards rapidly after 05:20~UT.

For the eruption from 9 May 2021, the filament in the quiet state was situated along the S-shaped magnetic PIL and represented sigmoid structures. In this case, again,  two clear eruption phases were observed: the slow-rising phase, which lasted between 09:30~UT and 10:00~UT, followed by a fast-rising phase, when the major part of the filament material erupted in the northeast direction. 
The eruption was accompanied by spreading ribbons at its base. This filament eruption disturbed the distant filament and that distant filament rose to a certain height and after that it fell back to the solar surface and was categorized as failed eruption.

Our observations provide evidence that the two filament eruptions were highly non-radial. \textls[-15]{The latitudinal deviation of the associated CMEs from the radial direction is ~analyzed.}

The event from 18 October 2017 was associated with CME, observed by the LASCO/C2 coronagraph. 
Our analysis showed a significant poleward latitudinal deviation of $29^\circ$ from the radial direction between the initial prominence location and the associated CME's loop. The non-radial motion stopped at about 15 $R_\odot$. 
This may be considered a relatively large deviation. 
For instance, \citep{cremade2004} found that the average deflection over the 1996–2002 period was $18.6^\circ$ based on their analysis of the non-radial motion of 276 CMEs. 
A deviation between $16^\circ$ and $28^\circ$ was reported by \citep{gui}, who examined the deflection of 10 CMEs in both latitude and longitude.
By examining 14 coronal mass ejections (CMEs) that were seen during Solar Cycle 23's decay and Cycle 24's rise, \citep{isavnin} found that the total longitudinal deflections varied between  $14^\circ$ and $28^\circ$ in absolute values.  
According to a recent statistical analysis of \citep{gregory} that examined over 14,000 CMEs from 1996 to 2022,  listed in the CDAW catalog, the greatest deflection values found in the field of view of LASCO coronagraphs were in the range of $\pm30^\circ$. 
The CME linked to the eruption of 9 May 2021, on the other hand, showed six-degree non-radial motion toward the equator that ended at roughly 2.25 $R_\odot$.

In the present study, we examined CME projected latitudinal deflection. Frequently non-radial motions are attributed to the presence of coronal holes (CHs). It was found that the CMEs typically drift  away from regions with open magnetic fields. CHs are areas of high Alfven speed that can result in CME deflections. Depending on how the CH and the CME source are positioned with respect to the observer, these deflections can cause a CME to move closer or farther from the Sun--Earth line \citep{gopal2}.
Many studies suggest that CMEs deviate away from nearby CHs to areas of lower magnetic energy (\citep{cremade,gopal2,kilpua,makela,sahade,cecere}).
Strong magnetic fields in the active region around the eruption or interactions between two CMEs could be also responsible for the near-Sun deflection of some CMEs, in addition to the influence of coronal holes (see \citep{kay,wang,gopal3}).

\begin{figure}[H]
	\includegraphics[width=8.5 cm]{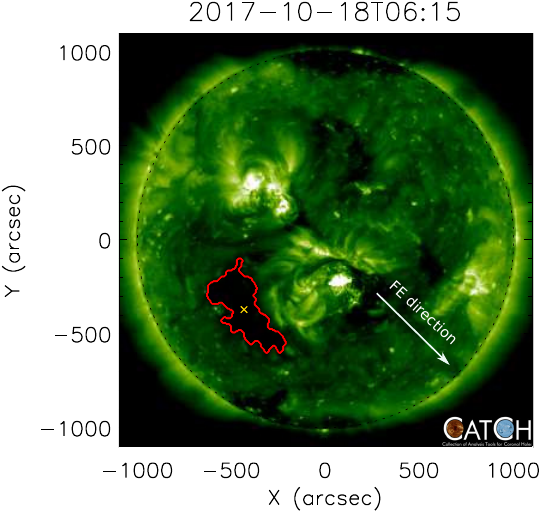}
	\caption{CH 
 boundary detection (red frame) for the event that took place on 18 October 2017. The direction of the eruption is shown by the white arrow.\label{dfig}}
\end{figure}

A low-latitude coronal hole was observed in EUVI/STEREO A 195 \AA\ images for the event, occurring on 2017 October 18. For CH boundary detection, we used the Collection of Analysis Tools for Coronal Holes (CATCH) \citep{heinemann} and the result is presented in Figure~\ref{dfig}.
Considering the absence of active regions near the eruption, we suggested that the CH influence was responsible for the observed CME deviation toward the pole.

\begin{figure}[H]
	\includegraphics[width=8 cm]{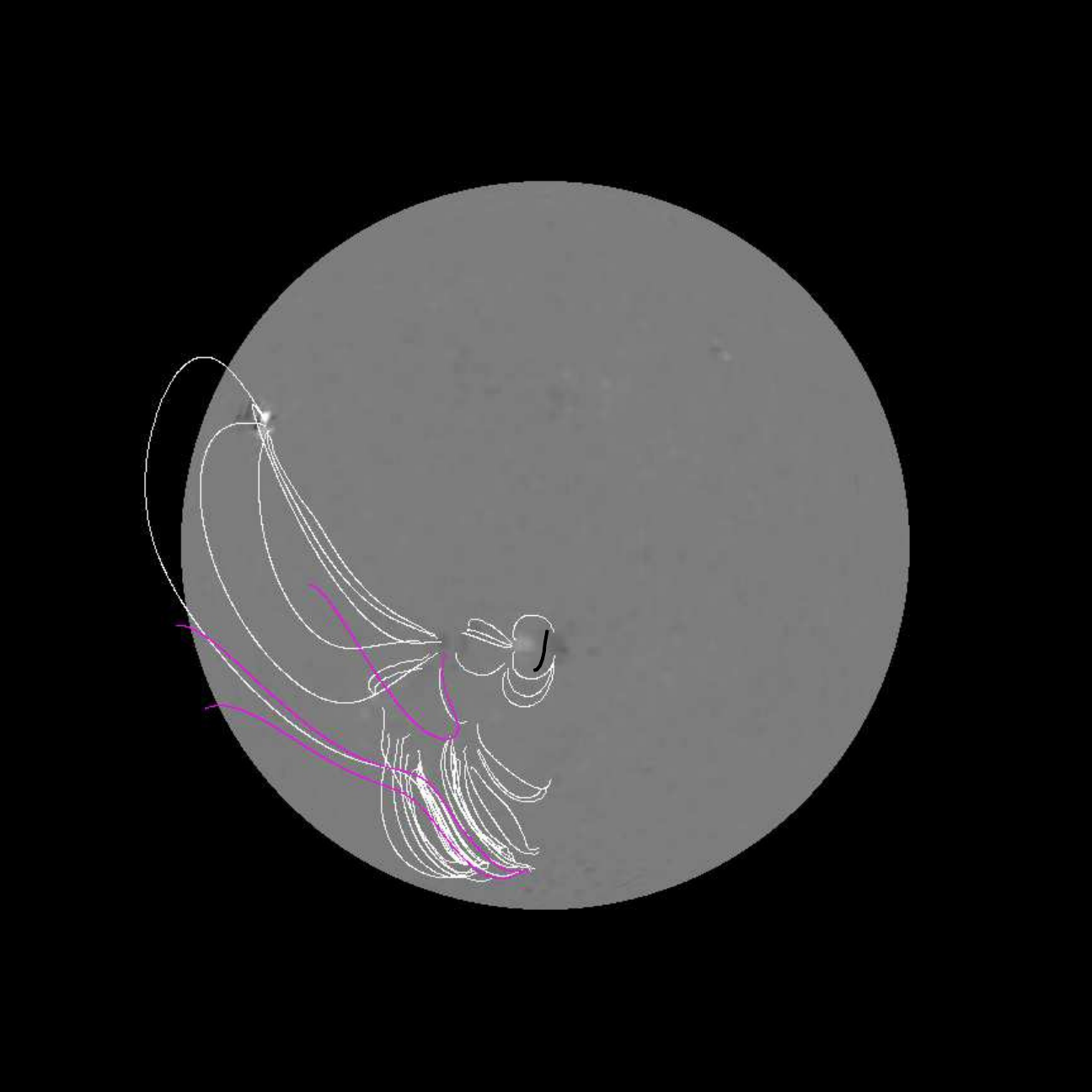}
	\caption{PFSS 
 extrapolation on 9 May 2021 at 10:59~UT. The open and closed field lines are represented in white and purple colors, respectively. The black line indicates the filament's initial position. 
 \label{fig_pfss}}
\end{figure}

We performed the Potential Field Source Surface (PFSS) extrapolation of the photospheric magnetic field for the 9 May 2021 event in order to explain the filament deflection and  the lack of observed associated CMEs in the SDO FOV. In Figure~\ref{fig_pfss} the PFSS extrapolation  at 10:59~UT is presented. 
The majority of erupted filament material  deviated eastward, probably channeled by the  open field lines, shown in purple in the figure.

\begin{figure}[H]
	\includegraphics[width=8 cm]{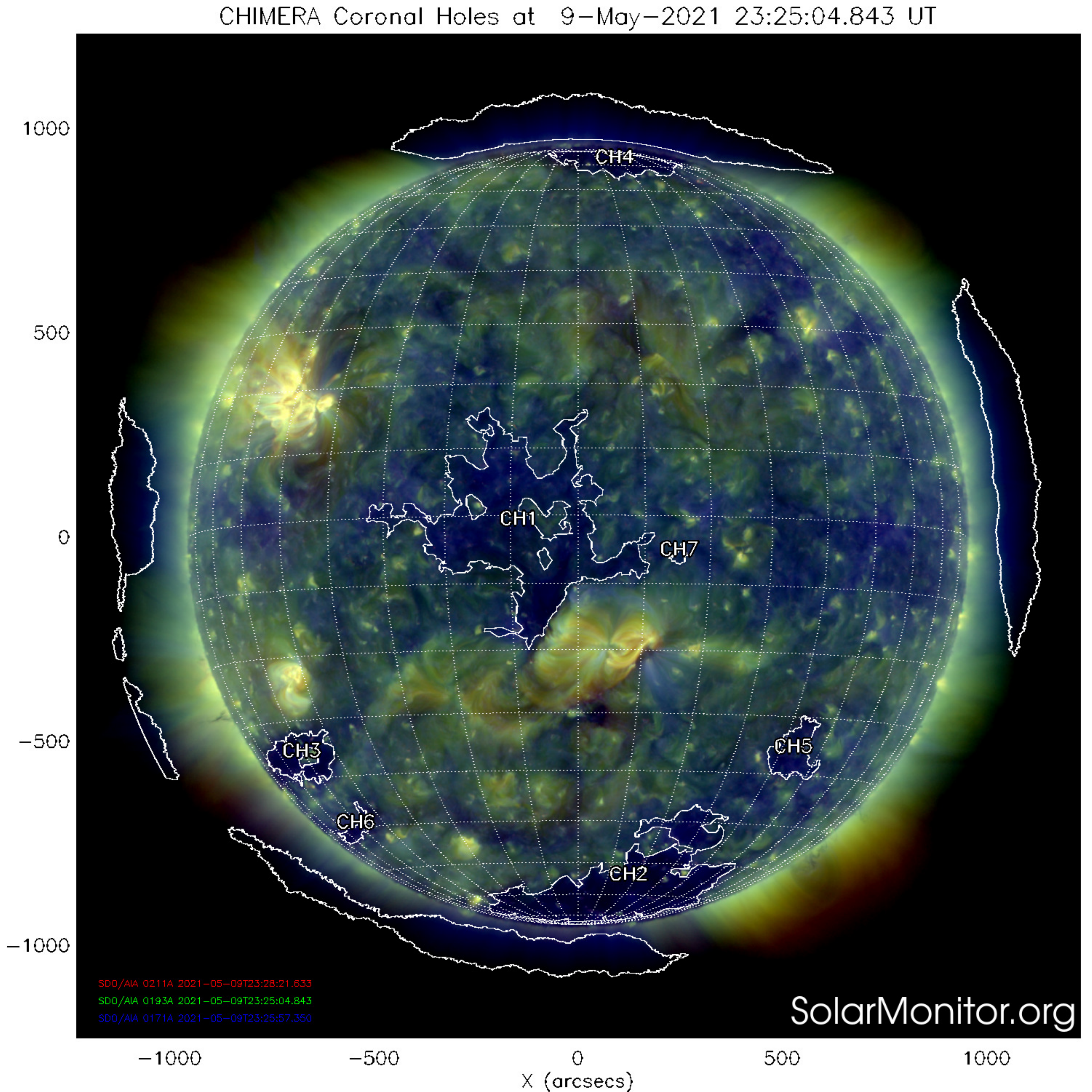}
	\caption{Image from Solar Monitor's coronal hole segmentation tool, showing the event from 9 May~2021. \label{ch_fig}}
\end{figure}

Solar Monitor's coronal hole segmentation tool (\url{https://www.solarmonitor.org/chimera.php}) showed a CH close to the southern pole for the 9 May 2021 eruption (\mbox{Figure~\ref{ch_fig}}), which could be the reason for the CME's deviation toward the equator. 
Furthermore, because of the presence of polar CHs, the CMEs tend to be deviated toward the equator during the Solar Cycle's rising phase.

\vspace{-9pt}

\vspace{-3pt}
\authorcontributions{Conceptualization,
 K.K. and R.C.; methodology, K.K., P.D. (Pooja Devi), and R.C.; software, M.D.; validation, all; formal analysis, R.C., K.K., and P.D. (Peter Duchlev) ; writing---original draft preparation, K.K., and P.D. (Peter Duchlev); writing---review and editing, all. All authors have read and agreed to the published version of the manuscript.}

\dataavailability{ The AIA images and spike files used in this article are publicly available from the JSOC at \url{jsoc.stanford.edu}.}

\acknowledgments{The authors acknowledge the open data policy of the SDO/AIA, SOHO/LASCO, STEREO/EUVI, and COR2 missions. 
 }

\conflictsofinterest{The authors declare no conflicts of interest.}

\begin{adjustwidth}{-\extralength}{0cm}
\reftitle{References}



\begin{thebibliography}{999}


\bibitem[Tandberg-Hanssen (1995)]{book1}
Tandberg-Hanssen, E. \textit{The Nature of Solar Prominences}; Astrophysics and Space Science Library; Kluwer Academic Publishers: Dordrecht, The Netherlands, 1995; Volume 199. 

\bibitem[Labrosse et al.(2010)]{labrosse}
Labrosse, N.; Heinzel, P.; Vial, J.-C.; Kucera, T.; Parenti, S.; Gunßr, S.; Schmieder, B.; Kilper, G. 
 Physics of solar prominences: I---Spectral diagnostics and non-LTE modelling. {\em SSRv} {\bf 2010}, {\em 243}, 243--332.

\bibitem[Mackay et al.(2010)]{macay}
Mackay, D.H.; Karpen, J.T.; Ballester, J.L.; Schmieder, B.; Aulanier, G. Physics of solar prominences: II---Magnetic structure and dynamics. {\em SSRv} {\bf 2010},
{\em 151}, 333--399.

\bibitem[Parenti (2014)]{parenti}
Parenti, S. Solar Prominences: Observations. {\em  Living Rev. Solar Phys.} {\bf 2014},
{\em 11}, 1--89.

\bibitem[Gibson (2018)]{gibson}
Gibson, S.E. Solar prominences: Theory and models. Fleshing out the magnetic skeleton. {\em Living Rev. Solar Phys.} {\bf 2018},
{\em 15}, 7--46.

\bibitem[Gilbert et al.(2007)]{gilbert}
Gilbert, H.R.; Alexander, D.; Liu, R. Filament Kinking and Its Implications for Eruption and Re-formation. {\em Sol. Phys.} {\bf 2007}, {\em 245}, 287--309.

\bibitem[Schrijver (2009)]{schr}
Schrijver, C.J. Driving major solar flares and eruptions: A review. {\em AdSpR} {\bf 2009},
{\em 43}, 739--755.

\bibitem[Sterling et al.(2007)]{st2007}
Sterling, A.; Harra, L.; Moore, R. New Evidence for the Role of Emerging Flux in a Solar Filament's Slow Rise Preceding Its CME-producing Fast Eruption. {\em ApJ} {\bf 2007}, {\em 669}, 1359--1371

\bibitem[Sterling et al.(2011)]{st2011}
Sterling, A.; Moore, R.; Freeland, S. Insights into Filament Eruption Onset from Solar Dynamics Observatory Observations. {\em ApJL} {\bf 2011}, {\em 731}, L3--L9.

\bibitem[Gopalswamy \& Thompson (2000)]{gopalTh}
Gopalswamy, N.; Thompson, B.J. Early life of coronal mass ejections. {\em JASTP} {\bf 2000}, {\em 62}, 1457--1469.

\bibitem[McCauley et al. (2015)]{mccauley}
McCauley, P.I.; Su, Y.N.; Schanche, N.; Evans, K.E.; Su, C.; McKillop, S.; Reeves, K.K.
 Prominence and Filament Eruptions Observed by the Solar Dynamics Observatory: Statistical Properties, Kinematics, and Online Catalog. {\em Sol. Phys.} {\bf 2015}, {\em 290}, 1703--1740.

\bibitem[Devi et al. (2021)]{devi}
Devi, P.; D\'{e}moulin, P.; Chandra, R.; Joshi, R.; Schmieder, B.; Joshi, B. Observations of a prominence eruption and loop contraction. {\em A\&A} {\bf 2021}, {\em 647}, A85--A97.

\bibitem[Williams et al. (2005)]{williams}
Williams, D.R.; T\"{o}r\"{o}k, T.; D\'{e}moulin, P.; van Driel-Gesztelyi, L.; Kliem, B. Eruption of a Kink-unstable Filament in NOAA Active Region 10696. {\em ApJ} {\bf 2005}, {\em 628}, L163--L166.

\bibitem[Chen \& Shibata (2000)]{chenshibata}
Chen, P.F.; Shibata, K. An Emerging Flux Trigger Mechanism for Coronal Mass Ejections. {\em ApJ} {\bf 2000}, {\em 545}, 524--531.

\bibitem[Munro et al.(1979)]{munro}
Munro, R.H.; Gosling, J.T.; Hildner, E.; MacQueen, R.M.; Pol, A.I.; Ross, C.L. The association of coronal mass ejection transients with other forms of solar activity. {\em Sol. Phys.} {\bf 1979}, {\em 61}, 201--215

\bibitem[St. Cyr \& Webb(1991)]{stcyr}
St Cyr, O.C.; Webb, D.F. Activity Associated with Coronal Mass Ejections at Solar Minimum - Solar Maximum Mission Observations from 1984--1986.  {\em Sol. Phys.} {\bf 1991}, {\em 136}, 379--394.

\bibitem[Lin et al. (2003)]{lin}
Lin, J.; Soon, W.; Baliunas, S.L. Theories of solar eruptions: A review.  {\em New Astron. Rev.} {\bf 2003}, {\em 47}, 53--84.

\bibitem[Choudhary \&Moore (2003)]{choudhary}
Choudhary, Debi Prasad; Moore, R.L. Filament eruption without coronal mass ejection. {\em GeoRL} {\bf 2003}, {\em 30}, 2107--2111.

\bibitem[Chandra et al.(2011)]{chandra}
Chandra, R.; Schmieder, B.; Mandrini, C.H.; Démoulin P; Schmieder B; Torok T; Chandra R. Homologous flares and magnetic field topology in active region NOAA 10501 on 20 november 2003. {\em Sol. Phys.} {\bf 2011}, {\em 269}, 83--104.

\bibitem[Chen (2011)]{chen}
Chen, P.F. Coronal mass ejections: Models and their observational basis. {\em Living Rev. Solar Phys.} {\bf 2011}, {\em 8}, 1--92.

\bibitem[Vourlidas et al.(2013)]{vourlidas}
Vourlidas, A.; Lynch, B. J.; Howard, R. A.; Li, Y. How many CMEs have flux ropes? Deciphering the signatures of shocks, flux ropes, and prominences in coronagraph observations of CMEs. {\em Sol. Phys.} {\bf 2013}, {\em 284}, 179--201

\bibitem[Schmieder et al.(2013)]{schmieder}
Schmieder, B.; Démoulin, P.; Aulanier, G. Solar filament eruptions and their physical role in triggering coronal mass ejections. {\em AdSpR} {\bf 2013}, {\em 51}, 1967--1980.

\bibitem[Gui et al.(2011)]{gui}
 Gui, B.; Shen, C.; Wang, Y.; Ye P.; Liu, J.; Wang, S.; Zhao, X. Quantitative Analysis of CME Deflections in the Corona. {\em Sol. Phys.} {\bf 2011}, {\em 271}, 111--139.

\bibitem[Isavnin et al.(2014)]{isavnin}
Isavnin, A.; Vourlidas, A.; Kilpua, E.K.J. Three-Dimensional Evolution of Flux-Rope CMEs and Its Relation to the Local Orientation of the Heliospheric Current Sheet. {\em Sol. Phys.} {\bf 2014}, {\em 298}, 2141--2156.

\bibitem[Lugaz et al.(2010)]{lugaz}
Lugaz, N. Accuracy and Limitations of Fitting and Stereoscopic Methods to Determine the Direction of Coronal Mass Ejections from Heliospheric Imagers Observations. {\em Sol. Phys.} {\bf 2010}, {\em 267}, 411--429.

\bibitem[Wang et al.(2014)]{wang1}
Wang, Y.; Wang, B.; Shen, C.; Shen, F.; Lugaz, N. Deflected propagation of a coronal mass ejection from the corona to interplanetary space. {\em JGRA} {\bf 2014}, {\em 119}, 5117--5132.
 
\bibitem[Lemen et al.(2012)]{lemen}
Lemen, J.R.; Title, A.M.; Akin, D.J.; Boerner, P.F.; Chou, C.; Drake, J.F.; Duncan, D.W.; Edwards, C.G.; Friedlaender, F.M.; \mbox{Heyman, G.F.; et al.} The Atmospheric Imaging Assembly (AIA) on the Solar Dynamics Observatory (SDO). {\em Sol. Phys.} {\bf 2012},
{\em 275}, 17--40.

\bibitem[Pesnell et al. (2012)]{pesnell}
Pesnell, W.D.; Thompson, B.J.; Chamberlin, P.C. The Solar Dynamics Observatory (SDO). {\em Sol. Phys.} {\bf 2012}, {\em 275}, 3--15.

\bibitem[Kaiser et al. (2008)]{kaiser}
Kaiser, M.L.; Kucera, T.A.; Davila, J.M.; St Cyr, O.C.; Guhathakurta, M.; Christian, E. The STEREO Mission: An Introduction. {\em SSRv} {\bf 2008}, {\em 136}, 5--16.

\bibitem[Thompson et al. (2003)]{thompson}
Thompson, W.T.; Davila, J.M.; Fisher, R.R.; Orwig, L.E.; Mentzell, J.E.; Hetherington, S.E.; Derro, R.J.; Federline, R.E.; Clark, D.C.; Chen, P.T.; et al. COR1 inner coronagraph for STEREO-SECCHI. In {\em Innovative Telescopes and Instrumentation for Solar Astrophysics. Proceedings of the SPIE}; Stephen L.K., Sergey, V.A., Eds.; Society of Photo-Optical Instrumentation Engineers (SPIE) Conference Series 
2003; Volume 4853, pp. 1--11.

\bibitem[Brueckner (1995)]{brueckner}
Brueckner, G.E. The Large Angle Spectroscopic Coronagraph (LASCO). {\em Sol. Phys.} {\bf 1995}, {\em 162}, 357--402

\bibitem[Yashiro et al. (2004)]{yashiro} 
Yashiro, S.; Gopalswamy, N.; Michalek, G.; St Cyr, O.C.; Plunkett, S.P.; Rich, N.B.; Howard, R.A. A catalog of white light coronal mass ejections observed by the SOHO spacecraft. {\em JGRA} {\bf 2004}, {\em 109}, A07105--A07116.

\bibitem[Gopalswamy et al. (2009)]{gopal} 
Gopalswamy, N.; Yashiro, S.; Michalek, G.; Stenborg G, Vourlidas A, Freeland S, Howard, R.
 The SOHO/LASCO CME Catalog. {\em  EM\&P} {\bf 2009}, {\em 104}, 295--313

\bibitem[Cheng et al. (2020)]{cheng}
Cheng, X.; Zhang, J.; Kliem, B.; Török, T.; Xing, C.; Zhou, Z.J.; Inhester, B.; Ding, M.D.
 Initiation and Early Kinematic Evolution of Solar Eruptions. {\em ApJ} {\bf 2020}, {\em 894}, 85--105.

\bibitem[Hood \& Priest (1981)]{hood}
 Hood, A.W.; Priest, E.R. Thermal Nonequilibrium---A Trigger for Solar Flares. {\em SoPh} {\bf 1981}, {\em 73}, 289--311.


\bibitem[Cremades and Bothmer (2004)]{cremade2004}
Cremades, H.; Bothmer,V. On the three-dimensional configuration of coronal mass ejections. {\em A\&A} {\bf 2004}, {\em 422}, 307--322.


\bibitem[Michalek et al. (2023)]{gregory}
Michalek, G.; Gopalswamy, N.; Yashiro, S.; Koleva, K. A Statistical Analysis of Deflection of Coronal Mass Ejections in the Field of View of LASCO Coronagraphs. {\em ApJ} {\bf 2023}, {\em 956}, 59--72.


\bibitem[Gopalswamy et al. (2009)]{gopal2} 
Gopalswamy, N.; M\"{a}kel\"{a}, P.; Xie, H.; Akiyama, S.; Yashiro, S. CME nteractions with coronal holes and their interplanetary consequences. {\em  JGRA} {\bf 2009}, {\em 114}, A00A22--A00A39.


\bibitem[Cremades et al. (2006)]{cremade}
Cremades, H.; Bothmer,V.; Tripathi, D. Properties of structured coronal mass ejections in solar cycle 23. {\em AdSpR} {\bf 2006}, {\em 38}, 461--465.


\bibitem[Kilpua et al. (2009)]{kilpua}
Kilpua, E.; Pomoell, J.; Vourlidas, A.; Vainio, R.; Luhmann, J.; Li, Y.; Schroeder, P.; Galvin, A.B.; Simunac, K. STEREO observations of interplanetary coronal mass ejections and prominence deflection during solar minimum period. {\em Ann. Geophys.} {\bf 2009}, {\em 27}, 4491--4503.

\bibitem[M\"{a}kel\"{a} et al. (2013)]{makela}
 M\"{a}kel\"{a}, P.; Gopalswamy, N.; Xie, H.; Mohamed, A.A.; Akiyama, S.; Yashiro, S. Coronal Hole Influence on the Observed Structure of Interplanetary CMEs. {\em SoPh} {\bf 2013}, {\em 284}, 59--75.

\bibitem[Sahade et al. (2020)]{sahade}
Sahade, A.; C\'{e}cere, M.; Krause, G. Influence of Coronal Holes on CME Deflections: Numerical Study. {\em ApJ} {\bf 2020}, {\em 896}, 53--64.

\bibitem[C\'{e}cere al. (2020)]{cecere}
C\'{e}cere, M.; Sieyra, M. V.; Cremades, H.; Mierla, M.; Sahade, A.; Stenborg, G.; Costa, A.; West, M.J.; D’Huys, E. Large non-radial propagation of a coronal mass ejection on 2011 January 24. {\em AdSpR} {\bf 2020}, {\em 65}, 1654--1662.

\bibitem[Kay et al. (2015)]{kay}
Kay, C.; Opher, M.; Evans, R.M. Global Trends of CME Deflections Based on CME and Solar Parameters. {\em ApJ} {\bf 2015}, {\em 805}, 168--188

\bibitem[Wang et al. (2015)]{wang}
Wang, R.; Liu, Ying D.; Dai, X.; Yang, Z.; Huang, C.; Hu, H. The Role of Active Region Coronal Magnetic Field in Determining Coronal Mass Ejection Propagation Direction. {\em ApJ} {\bf 2015}, {\em 814}, 80--89.

\bibitem[Gopalswamy et al. (2001)]{gopal3} 
Gopalswamy, N.; Yashiro, S.; Kaiser, M.L.; Howard, R.A.; Bougeret, J.L. Radio Signatures of Coronal Mass Ejection Interaction: Coronal Mass Ejection Cannibalism? {\em  ApJ} {\bf 2001}, {\em 584}, 91--94.

\bibitem[Heinemann et al. (2019)]{heinemann}
Heinemann, S. G., Temmer, M., Heinemann, N.; Dissauer, K.; Samara, E.; Jerčić, V.; Hofmeister, S.J.; Veronig, A.M. Statistical Analysis and Catalog of Non-polar Coronal Holes Covering the SDO-Era Using CATCH. {\em SoPh} {\bf 2019}, {\em 294}, 144--168.
\end{thebibliography}


\isAPAandChicago{}{%

}
\PublishersNote{}
\end{adjustwidth}
\end{document}